\documentstyle[11pt,aaspp4]{article}
\newcommand{\etal}{et~al.}

\def\PsfigVersion{1.10}
\def\setDriver{\DvipsDriver} % \DvipsDriver or \OzTeXDriver
\ifx\undefined\psfig\else \fi
\let\LaTeXAtSign=\@
\let\@=\relax
\edef\psfigRestoreAt{\catcode`\@=\number\catcode`@\relax}
\catcode`\@=11\relax
\newwrite\@unused
\def\ps@typeout#1{{\let\protect\string\immediate\write\@unused{#1}}}

\def\DvipsDriver{
	\ps@typeout{psfig/tex \PsfigVersion -dvips}
\def\PsfigSpecials{\DvipsSpecials} 	\def\ps@dir{/}
\def\ps@predir{} }
\def\OzTeXDriver{
	\ps@typeout{psfig/tex \PsfigVersion -oztex}
	\def\PsfigSpecials{\OzTeXSpecials}
	\def\ps@dir{:}
	\def\ps@predir{:}
	\catcode`\^^J=5
}

\def\figurepath{./:}

\def\DoPaths#1{\expandafter\EachPath#1\stoplist}
\def\leer{}
\def\EachPath#1:#2\stoplist{% #1 part of the list (delimiter :)
  \ExistsFile{#1}{\SearchedFile}
  \ifx#2\leer
  \else
    \expandafter\EachPath#2\stoplist
  \fi}
\def\ps@dir{/}
\def\ExistsFile#1#2{%
   \openin1=\ps@predir#1\ps@dir#2
   \ifeof1
       \closein1
   \else
       \closein1
        \ifx\ps@founddir\leer
           \edef\ps@founddir{#1}
        \fi
   \fi}
\def\get@dir#1{%
  \def\ps@founddir{}
  \def\SearchedFile{#1}
  \DoPaths\figurepath
}
\def\@nnil{\@nil}
\def\@empty{}
\def\@psdonoop#1\@@#2#3{}
\def\@psdo#1:=#2\do#3{\edef\@psdotmp{#2}\ifx\@psdotmp\@empty \else
    \expandafter\@psdoloop#2,\@nil,\@nil\@@#1{#3}\fi}
\def\@psdoloop#1,#2,#3\@@#4#5{\def#4{#1}\ifx #4\@nnil \else
       #5\def#4{#2}\ifx #4\@nnil \else#5\@ipsdoloop #3\@@#4{#5}\fi\fi}
\def\@ipsdoloop#1,#2\@@#3#4{\def#3{#1}\ifx #3\@nnil 
       \let\@nextwhile=\@psdonoop \else
      #4\relax\let\@nextwhile=\@ipsdoloop\fi\@nextwhile#2\@@#3{#4}}
\def\@tpsdo#1:=#2\do#3{\xdef\@psdotmp{#2}\ifx\@psdotmp\@empty \else
    \@tpsdoloop#2\@nil\@nil\@@#1{#3}\fi}
\def\@tpsdoloop#1#2\@@#3#4{\def#3{#1}\ifx #3\@nnil 
       \let\@nextwhile=\@psdonoop \else
      #4\relax\let\@nextwhile=\@tpsdoloop\fi\@nextwhile#2\@@#3{#4}}
\ifx\undefined\fbox
\newdimen\fboxrule
\newdimen\fboxsep
\newdimen\ps@tempdima
\newbox\ps@tempboxa
\long\def\fbox#1{\leavevmode\setbox\ps@tempboxa\hbox{#1}\ps@tempdima\fboxrule
    \advance\ps@tempdima \fboxsep \advance\ps@tempdima \dp\ps@tempboxa
   \hbox{\lower \ps@tempdima\hbox
  {\vbox{\hrule height \fboxrule
          \hbox{\vrule width \fboxrule \hskip\fboxsep
          \vbox{\vskip\fboxsep \box\ps@tempboxa\vskip\fboxsep}\hskip 
                 \fboxsep\vrule width \fboxrule}
                 \hrule height \fboxrule}}}}
\fi
\newread\ps@stream
\newif\ifnot@eof       % continue looking for the bounding box?
\newif\if@noisy        % report what you're making?
\newif\if@atend        % %%BoundingBox: has (at end) specification
\newif\if@psfile       % does this look like a PostScript file?
{\catcode`\%=12\global\gdef\epsf@start{%!}}
\def\epsf@PS{PS}
\def\epsf@getbb#1{%
\openin\ps@stream=\ps@predir#1
\ifeof\ps@stream\ps@typeout{Error, File #1 not found}\else
   {\not@eoftrue \chardef\other=12
    \def\do##1{\catcode`##1=\other}\dospecials \catcode`\ =10
    \loop
       \if@psfile
	  \read\ps@stream to \epsf@fileline
       \else{
	  \obeyspaces
          \read\ps@stream to \epsf@tmp\global\let\epsf@fileline\epsf@tmp}
       \fi
       \ifeof\ps@stream\not@eoffalse\else
       \if@psfile\else
       \expandafter\epsf@test\epsf@fileline:. \\%
       \fi
          \expandafter\epsf@aux\epsf@fileline:. \\%
       \fi
   \ifnot@eof\repeat
   }\closein\ps@stream\fi}%
\long\def\epsf@test#1#2#3:#4\\{\def\epsf@testit{#1#2}
			\ifx\epsf@testit\epsf@start\else
\ps@typeout{Warning! File does not start with `\epsf@start'.  It may not be a PostScript file.}
			\fi
			\@psfiletrue} % don't test after 1st line
{\catcode`\%=12\global\let\epsf@percent=%\global\def\epsf@bblit{%BoundingBox}}
\long\def\epsf@aux#1#2:#3\\{\ifx#1\epsf@percent
   \def\epsf@testit{#2}\ifx\epsf@testit\epsf@bblit
	\@atendfalse
        \epsf@atend #3 . \\%
	\if@atend	
	   \if@verbose{
		\ps@typeout{psfig: found `(atend)'; continuing search}
	   }\fi
        \else
        \epsf@grab #3 . . . \\%
        \not@eoffalse
        \global\no@bbfalse
        \fi
   \fi\fi}%
\def\epsf@grab #1 #2 #3 #4 #5\\{%
   \global\def\epsf@llx{#1}\ifx\epsf@llx\empty
      \epsf@grab #2 #3 #4 #5 .\\\else
   \global\def\epsf@lly{#2}%
   \global\def\epsf@urx{#3}\global\def\epsf@ury{#4}\fi}%
\def\epsf@atendlit{(atend)} 
\def\epsf@atend #1 #2 #3\\{%
   \def\epsf@tmp{#1}\ifx\epsf@tmp\empty
      \epsf@atend #2 #3 .\\\else
   \ifx\epsf@tmp\epsf@atendlit\@atendtrue\fi\fi}

\chardef\psletter = 11 % won't conflict with \begin{letter} now...
\chardef\other = 12

\newif \ifdebug %%% turn me on to see TeX hard at work ...
\newif\ifc@mpute %%% don't need to compute some values
\c@mputetrue % but assume that we do

\let\then = \relax
\def\r@dian{pt }
\let\r@dians = \r@dian
\let\dimensionless@nit = \r@dian
\let\dimensionless@nits = \dimensionless@nit
\def\internal@nit{sp }
\let\internal@nits = \internal@nit
\newif\ifstillc@nverging
\def \Mess@ge #1{\ifdebug \then \message {#1} \fi}

{ %%% Things that need abnormal catcodes %%%
	\catcode `\@ = \psletter
	\gdef \nodimen {\expandafter \n@dimen \the \dimen}
	\gdef \term #1 #2 #3%
	       {\edef \t@ {\the #1}%%% freeze parameter 1 (count, by value)
		\edef \t@@ {\expandafter \n@dimen \the #2\r@dian}%
		\t@rm {\t@} {\t@@} {#3}%
	       }
	\gdef \t@rm #1 #2 #3%
	       {{%
		\count 0 = 0
		\dimen 0 = 1 \dimensionless@nit
		\dimen 2 = #2\relax
		\Mess@ge {Calculating term #1 of \nodimen 2}%
		\loop
		\ifnum	\count 0 < #1
		\then	\advance \count 0 by 1
			\Mess@ge {Iteration \the \count 0 \space}%
			\Multiply \dimen 0 by {\dimen 2}%
			\Mess@ge {After multiplication, term = \nodimen 0}%
			\Divide \dimen 0 by {\count 0}%
			\Mess@ge {After division, term = \nodimen 0}%
		\repeat
		\Mess@ge {Final value for term #1 of 
				\nodimen 2 \space is \nodimen 0}%
		\xdef \Term {#3 = \nodimen 0 \r@dians}%
		\aftergroup \Term
	       }}
	\catcode `\p = \other
	\catcode `\t = \other
	\gdef \n@dimen #1pt{#1} %%% throw away the ``pt''
}

\def \Divide #1by #2{\divide #1 by #2} %%% just a synonym

\def \Multiply #1by #2%%% allows division of a dimen by a dimen
       {{%%% should really freeze parameter 2 (dimen, passed by value)
	\count 0 = #1\relax
	\count 2 = #2\relax
	\count 4 = 65536
	\Mess@ge {Before scaling, count 0 = \the \count 0 \space and
			count 2 = \the \count 2}%
	\ifnum	\count 0 > 32767 %%% do our best to avoid overflow
	\then	\divide \count 0 by 4
		\divide \count 4 by 4
	\else	\ifnum	\count 0 < -32767
		\then	\divide \count 0 by 4
			\divide \count 4 by 4
		\else
		\fi
	\fi
	\ifnum	\count 2 > 32767 %%% while retaining reasonable accuracy
	\then	\divide \count 2 by 4
		\divide \count 4 by 4
	\else	\ifnum	\count 2 < -32767
		\then	\divide \count 2 by 4
			\divide \count 4 by 4
		\else
		\fi
	\fi
	\multiply \count 0 by \count 2
	\divide \count 0 by \count 4
	\xdef \product {#1 = \the \count 0 \internal@nits}%
	\aftergroup \product
       }}

\def\r@duce{\ifdim\dimen0 > 90\r@dian \then   % sin(x+90) = sin(180-x)
		\multiply\dimen0 by -1
		\advance\dimen0 by 180\r@dian
		\r@duce
	    \else \ifdim\dimen0 < -90\r@dian \then  % sin(-x) = sin(360+x)
		\advance\dimen0 by 360\r@dian
		\r@duce
		\fi
	    \fi}

\def\Sine#1%
       {{%
	\dimen 0 = #1 \r@dian
	\r@duce
	\ifdim\dimen0 = -90\r@dian \then
	   \dimen4 = -1\r@dian
	   \c@mputefalse
	\fi
	\ifdim\dimen0 = 90\r@dian \then
	   \dimen4 = 1\r@dian
	   \c@mputefalse
	\fi
	\ifdim\dimen0 = 0\r@dian \then
	   \dimen4 = 0\r@dian
	   \c@mputefalse
	\fi
	\ifc@mpute \then
		\divide\dimen0 by 180
		\dimen0=3.141592654\dimen0
		\dimen 2 = 3.1415926535897963\r@dian %%% a well-known constant
		\divide\dimen 2 by 2 %%% we only deal with -pi/2 : pi/2
		\Mess@ge {Sin: calculating Sin of \nodimen 0}%
		\count 0 = 1 %%% see power-series expansion for sine
		\dimen 2 = 1 \r@dian %%% ditto
		\dimen 4 = 0 \r@dian %%% ditto
		\loop
			\ifnum	\dimen 2 = 0 %%% then we've done
			\then	\stillc@nvergingfalse 
			\else	\stillc@nvergingtrue
			\fi
			\ifstillc@nverging %%% then calculate next term
			\then	\term {\count 0} {\dimen 0} {\dimen 2}%
				\advance \count 0 by 2
				\count 2 = \count 0
				\divide \count 2 by 2
				\ifodd	\count 2 %%% signs alternate
				\then	\advance \dimen 4 by \dimen 2
				\else	\advance \dimen 4 by -\dimen 2
				\fi
		\repeat
	\fi		
			\xdef \sine {\nodimen 4}%
       }}

\def\Cosine#1{\ifx\sine\UnDefined\edef\Savesine{\relax}\else
		             \edef\Savesine{\sine}\fi
	{\dimen0=#1\r@dian\advance\dimen0 by 90\r@dian
	 \Sine{\nodimen 0}
	 \xdef\cosine{\sine}
	 \xdef\sine{\Savesine}}}	      
\def\psdraft{
	\def\@psdraft{0}
}
\def\psfull{
	\def\@psdraft{100}
}

\psfull

\newif\if@scalefirst
\def\psscalefirst{\@scalefirsttrue}
\def\psrotatefirst{\@scalefirstfalse}
\psrotatefirst

\newif\if@draftbox
\def\psnodraftbox{
	\@draftboxfalse
}
\def\psdraftbox{
	\@draftboxtrue
}
\@draftboxtrue

\newif\if@prologfile
\newif\if@postlogfile
\def\pssilent{
	\@noisyfalse
}
\def\psnoisy{
	\@noisytrue
}
\psnoisy
\newif\if@bbllx
\newif\if@bblly
\newif\if@bburx
\newif\if@bbury
\newif\if@height
\newif\if@width
\newif\if@rheight
\newif\if@rwidth
\newif\if@angle
\newif\if@clip
\newif\if@verbose
\def\@p@@sclip#1{\@cliptrue}
\newif\if@decmpr
\def\@p@@sfigure#1{\def\@p@sfile{null}\def\@p@sbbfile{null}\@decmprfalse
   \openin1=\ps@predir#1
   \ifeof1
	\closein1
	\get@dir{#1}
	\ifx\ps@founddir\leer
		\openin1=\ps@predir#1.bb
		\ifeof1
			\closein1
			\get@dir{#1.bb}
			\ifx\ps@founddir\leer
				\ps@typeout{Can't find #1 in \figurepath}
			\else
				\@decmprtrue
				\def\@p@sfile{\ps@founddir\ps@dir#1}
				\def\@p@sbbfile{\ps@founddir\ps@dir#1.bb}
			\fi
		\else
			\closein1
			\@decmprtrue
			\def\@p@sfile{#1}
			\def\@p@sbbfile{#1.bb}
		\fi
	\else
		\def\@p@sfile{\ps@founddir\ps@dir#1}
		\def\@p@sbbfile{\ps@founddir\ps@dir#1}
	\fi
   \else
	\closein1
	\def\@p@sfile{#1}
	\def\@p@sbbfile{#1}
   \fi
}
\def\@p@@sfile#1{\@p@@sfigure{#1}}
\def\@p@@sbbllx#1{
		\@bbllxtrue
		\dimen100=#1
		\edef\@p@sbbllx{\number\dimen100}
}
\def\@p@@sbblly#1{
		\@bbllytrue
		\dimen100=#1
		\edef\@p@sbblly{\number\dimen100}
}
\def\@p@@sbburx#1{
		\@bburxtrue
		\dimen100=#1
		\edef\@p@sbburx{\number\dimen100}
}
\def\@p@@sbbury#1{
		\@bburytrue
		\dimen100=#1
		\edef\@p@sbbury{\number\dimen100}
}
\def\@p@@sheight#1{
		\@heighttrue
		\dimen100=#1
   		\edef\@p@sheight{\number\dimen100}
}
\def\@p@@swidth#1{
		\@widthtrue
		\dimen100=#1
		\edef\@p@swidth{\number\dimen100}
}
\def\@p@@srheight#1{
		\@rheighttrue
		\dimen100=#1
		\edef\@p@srheight{\number\dimen100}
}
\def\@p@@srwidth#1{
		\@rwidthtrue
		\dimen100=#1
		\edef\@p@srwidth{\number\dimen100}
}
\def\@p@@sangle#1{
		\@angletrue
		\edef\@p@sangle{#1} %\number\dimen100}
}
\def\@p@@ssilent#1{ 
		\@verbosefalse
}
\def\@p@@sprolog#1{\@prologfiletrue\def\@prologfileval{#1}}
\def\@p@@spostlog#1{\@postlogfiletrue\def\@postlogfileval{#1}}
\def\@cs@name#1{\csname #1\endcsname}
\def\@setparms#1=#2,{\@cs@name{@p@@s#1}{#2}}
\def\ps@init@parms{
		\@bbllxfalse \@bbllyfalse
		\@bburxfalse \@bburyfalse
		\@heightfalse \@widthfalse
		\@rheightfalse \@rwidthfalse
		\def\@p@sbbllx{}\def\@p@sbblly{}
		\def\@p@sbburx{}\def\@p@sbbury{}
		\def\@p@sheight{}\def\@p@swidth{}
		\def\@p@srheight{}\def\@p@srwidth{}
		\def\@p@sangle{0}
		\def\@p@sfile{} \def\@p@sbbfile{}
		\def\@p@scost{10}
		\def\@sc{}
		\@prologfilefalse
		\@postlogfilefalse
		\@clipfalse
		\if@noisy
			\@verbosetrue
		\else
			\@verbosefalse
		\fi
}
\def\parse@ps@parms#1{
	 	\@psdo\@psfiga:=#1\do
		   {\expandafter\@setparms\@psfiga,}}
\newif\ifno@bb
\def\bb@missing{
	\if@verbose{
		\ps@typeout{psfig: searching \@p@sbbfile \space  for bounding box}
	}\fi
	\no@bbtrue
	\epsf@getbb{\@p@sbbfile}
        \ifno@bb \else \bb@cull\epsf@llx\epsf@lly\epsf@urx\epsf@ury\fi
}	
\def\bb@cull#1#2#3#4{
	\dimen100=#1 bp\edef\@p@sbbllx{\number\dimen100}
	\dimen100=#2 bp\edef\@p@sbblly{\number\dimen100}
	\dimen100=#3 bp\edef\@p@sbburx{\number\dimen100}
	\dimen100=#4 bp\edef\@p@sbbury{\number\dimen100}
	\no@bbfalse
}
\newdimen\p@intvaluex
\newdimen\p@intvaluey
\def\rotate@#1#2{{\dimen0=#1 sp\dimen1=#2 sp
		  \global\p@intvaluex=\cosine\dimen0
		  \dimen3=\sine\dimen1
		  \global\advance\p@intvaluex by -\dimen3
		  \global\p@intvaluey=\sine\dimen0
		  \dimen3=\cosine\dimen1
		  \global\advance\p@intvaluey by \dimen3
		  }}
\def\compute@bb{
		\no@bbfalse
		\if@bbllx \else \no@bbtrue \fi
		\if@bblly \else \no@bbtrue \fi
		\if@bburx \else \no@bbtrue \fi
		\if@bbury \else \no@bbtrue \fi
		\ifno@bb \bb@missing \fi
		\ifno@bb \ps@typeout{FATAL ERROR: no bb supplied or found}
			\no-bb-error
		\fi
		\count203=\@p@sbburx
		\count204=\@p@sbbury
		\advance\count203 by -\@p@sbbllx
		\advance\count204 by -\@p@sbblly
		\edef\ps@bbw{\number\count203}
		\edef\ps@bbh{\number\count204}
		\if@angle 
			\Sine{\@p@sangle}\Cosine{\@p@sangle}
	        	{\dimen100=\maxdimen\xdef\r@p@sbbllx{\number\dimen100}
					    \xdef\r@p@sbblly{\number\dimen100}
			                    \xdef\r@p@sbburx{-\number\dimen100}
					    \xdef\r@p@sbbury{-\number\dimen100}}
                        \def\minmaxtest{
			   \ifnum\number\p@intvaluex<\r@p@sbbllx
			      \xdef\r@p@sbbllx{\number\p@intvaluex}\fi
			   \ifnum\number\p@intvaluex>\r@p@sbburx
			      \xdef\r@p@sbburx{\number\p@intvaluex}\fi
			   \ifnum\number\p@intvaluey<\r@p@sbblly
			      \xdef\r@p@sbblly{\number\p@intvaluey}\fi
			   \ifnum\number\p@intvaluey>\r@p@sbbury
			      \xdef\r@p@sbbury{\number\p@intvaluey}\fi
			   }
			\rotate@{\@p@sbbllx}{\@p@sbblly}
			\minmaxtest
			\rotate@{\@p@sbbllx}{\@p@sbbury}
			\minmaxtest
			\rotate@{\@p@sbburx}{\@p@sbblly}
			\minmaxtest
			\rotate@{\@p@sbburx}{\@p@sbbury}
			\minmaxtest
			\edef\@p@sbbllx{\r@p@sbbllx}\edef\@p@sbblly{\r@p@sbblly}
			\edef\@p@sbburx{\r@p@sbburx}\edef\@p@sbbury{\r@p@sbbury}
		\fi
		\count203=\@p@sbburx
		\count204=\@p@sbbury
		\advance\count203 by -\@p@sbbllx
		\advance\count204 by -\@p@sbblly
		\edef\@bbw{\number\count203}
		\edef\@bbh{\number\count204}
}
\def\in@hundreds#1#2#3{\count240=#2 \count241=#3
		     \count100=\count240	% 100 is first digit #2/#3
		     \divide\count100 by \count241
		     \count101=\count100
		     \multiply\count101 by \count241
		     \advance\count240 by -\count101
		     \multiply\count240 by 10
		     \count101=\count240	%101 is second digit of #2/#3
		     \divide\count101 by \count241
		     \count102=\count101
		     \multiply\count102 by \count241
		     \advance\count240 by -\count102
		     \multiply\count240 by 10
		     \count102=\count240	% 102 is the third digit
		     \divide\count102 by \count241
		     \count200=#1\count205=0
		     \count201=\count200
			\multiply\count201 by \count100
		 	\advance\count205 by \count201
		     \count201=\count200
			\divide\count201 by 10
			\multiply\count201 by \count101
			\advance\count205 by \count201
		     \count201=\count200
			\divide\count201 by 100
			\multiply\count201 by \count102
			\advance\count205 by \count201
		     \edef\@result{\number\count205}
}
\def\compute@wfromh{
		\in@hundreds{\@p@sheight}{\@bbw}{\@bbh}
		\edef\@p@swidth{\@result}
}
\def\compute@hfromw{
	        \in@hundreds{\@p@swidth}{\@bbh}{\@bbw}
		\edef\@p@sheight{\@result}
}
\def\compute@handw{
		\if@height 
			\if@width
			\else
				\compute@wfromh
			\fi
		\else 
			\if@width
				\compute@hfromw
			\else
				\edef\@p@sheight{\@bbh}
				\edef\@p@swidth{\@bbw}
			\fi
		\fi
}
\def\compute@resv{
		\if@rheight \else \edef\@p@srheight{\@p@sheight} \fi
		\if@rwidth \else \edef\@p@srwidth{\@p@swidth} \fi
}
\def\compute@sizes{
	\compute@bb
	\if@scalefirst\if@angle
	\if@width
	   \in@hundreds{\@p@swidth}{\@bbw}{\ps@bbw}
	   \edef\@p@swidth{\@result}
	\fi
	\if@height
	   \in@hundreds{\@p@sheight}{\@bbh}{\ps@bbh}
	   \edef\@p@sheight{\@result}
	\fi
	\fi\fi
	\compute@handw
	\compute@resv}
\def\OzTeXSpecials{
	\special{empty.ps /@isp {true} def}
	\special{empty.ps \@p@swidth \space \@p@sheight \space
			\@p@sbbllx \space \@p@sbblly \space
			\@p@sbburx \space \@p@sbbury \space
			startTexFig \space }
	\if@clip{
		\if@verbose{
			\ps@typeout{(clip)}
		}\fi
		\special{empty.ps doclip \space }
	}\fi
	\if@angle{
		\if@verbose{
			\ps@typeout{(rotate)}
		}\fi
		\special {empty.ps \@p@sangle \space rotate \space} 
	}\fi
	\if@prologfile
	    \special{\@prologfileval \space } \fi
	\if@decmpr{
		\if@verbose{
			\ps@typeout{psfig: Compression not available
			in OzTeX version \space }
		}\fi
	}\else{
		\if@verbose{
			\ps@typeout{psfig: including \@p@sfile \space }
		}\fi
		\special{epsf=\ps@predir\@p@sfile \space }
	}\fi
	\if@postlogfile
	    \special{\@postlogfileval \space } \fi
	\special{empty.ps /@isp {false} def}
}
\def\DvipsSpecials{
	\special{ps::[begin] 	\@p@swidth \space \@p@sheight \space
			\@p@sbbllx \space \@p@sbblly \space
			\@p@sbburx \space \@p@sbbury \space
			startTexFig \space }
	\if@clip{
		\if@verbose{
			\ps@typeout{(clip)}
		}\fi
		\special{ps:: doclip \space }
	}\fi
	\if@angle
		\if@verbose{
			\ps@typeout{(clip)}
		}\fi
		\special {ps:: \@p@sangle \space rotate \space} 
	\fi
	\if@prologfile
	    \special{ps: plotfile \@prologfileval \space } \fi
	\if@decmpr{
		\if@verbose{
			\ps@typeout{psfig: including \@p@sfile.Z \space }
		}\fi
		\special{ps: plotfile "`zcat \@p@sfile.Z" \space }
	}\else{
		\if@verbose{
			\ps@typeout{psfig: including \@p@sfile \space }
		}\fi
		\special{ps: plotfile \@p@sfile \space }
	}\fi
	\if@postlogfile
	    \special{ps: plotfile \@postlogfileval \space } \fi
	\special{ps::[end] endTexFig \space }
}
\def\psfig#1{\vbox {
	\ps@init@parms
	\parse@ps@parms{#1}
	\compute@sizes
	\ifnum\@p@scost<\@psdraft{
		\PsfigSpecials 
		\vbox to \@p@srheight sp{
			\hbox to \@p@srwidth sp{
				\hss
			}
		\vss
		}
	}\else{
		\if@draftbox{		
			\hbox{\fbox{\vbox to \@p@srheight sp{
			\vss
			\hbox to \@p@srwidth sp{ \hss 
			 \hss }
			\vss
			}}}
		}\else{
			\vbox to \@p@srheight sp{
			\vss
			\hbox to \@p@srwidth sp{\hss}
			\vss
			}
		}\fi

	}\fi
}}
\psfigRestoreAt
\setDriver
\let\@=\LaTeXAtSign

\def\gtorder{\mathrel{\raise.3ex\hbox{$>$}\mkern-14mu
             \lower0.6ex\hbox{$\sim$}}}
\def\ltorder{\mathrel{\raise.3ex\hbox{$<$}\mkern-14mu
             \lower0.6ex\hbox{$\sim$}}}

\begin{document}

\title{Cusped Mass Models Of Gravitational Lenses}

\author{J.A. Mu\~noz$^{1}$, C.S. Kochanek$^{2}$ and C.R. Keeton$^{3}$}

\affil{$^{1}$Instituto de Astrof\'{\i}sica de Canarias, E-38200 La Laguna, Tenerife, Spain}
\affil{email: jmunoz@ll.iac.es}
\affil{$^{2}$Harvard-Smithsonian Center for Astrophysics, 60 Garden St.,
  Cambridge, MA 02138}
\affil{email: ckochanek@cfa.harvard.edu}
\affil{$^{3}$Steward Observatory, University of Arizona, Tucson, AZ 85721}
\affil{email: ckeeton@as.arizona.edu}

\begin{abstract}
Recent observations of galaxy luminosity profiles and dark matter simulations
find luminosity and mass distributions characterized by central cusps 
rather than finite core radii.  We introduce and implement a set of cusped 
ellipsoidal lens models which include limits similar to the Jaffe, Hernquist,
$\eta$ and NFW models and apply them to the gravitational lenses 
APM~08279+5255 and B~1933+503.  A successful model of APM~08279+5255
with its central, odd image requires a very shallow cusp, $\gamma \ltorder 0.4$
where $\rho \propto r^{-\gamma}$ as $r\rightarrow 0$, which is similar
to a core rather than the favored $1 \ltorder \gamma \ltorder 2$ cusps.  
B~1933+503, by contrast, is
well modeled with a steep density cusp, $1.6 \ltorder \gamma \ltorder 2.0$. 
\end{abstract}


\section{Introduction}

The families of density distributions used to model gravitational lenses are 
rooted in the lore of the early 1980s, where massive galaxies were believed to have finite
core radii (e.g. Lauer 1985).  Hence, surveys of lens models almost always use 
models related to the
softened power-law surface density, $\Sigma \propto (m^2+s^2)^{(\alpha-2)/2}$ where
$m^2$ is a two-dimensional ellipsoidal coordinate and $s$ is the core radius, 
as a template for exploring the effects of the radial mass distribution on 
lensing phenomenon (e.g. Kochanek 1995, Chen, Kochanek \& Hewitt 1995, Grogin \& Narayan 1996, 
Chae, Turnshek \& Khersonsky 1998, Schmidt, Webster \& Lewis 1998, Keeton \& Kochanek 1998,
Chae 1999, Keeton et al. 2000, Cohn et al. 2001). 
The asymptotic exponent $\alpha$ covers a range of historical models for galaxies from 
the singular isothermal ellipsoid (or SIE, $\alpha=1$ and $s=0$), to the modified
Hubble profile ($\alpha=0$) and the Plummer model ($\alpha=-2$). Viewed as 
projections of thin disks, these models include the Mestel (1963) disk 
(equivalent to the SIE), and the Kuzmin (1956) disk ($\alpha=-1$).
For $\alpha=-1$ and $1$ there are simple analytic expressions for the 
potentials (see Keeton \& Kochanek 1998 and references therein), and relatively
fast approximation schemes have been developed for a broad range of exponents
(Barkana 1998, Chae, Khersonsky \& Turnshek 1998).  Unfortunately, the only 
model from this broad class of potentials that can be regarded as 
astrophysical is the SIE.

Since the advent of HST and high resolution numerical simulations, our view of 
the central regions of galaxies has changed significantly, in that massive
galaxies have central
cusps rather than central core radii in both high resolution observations of 
luminosity profiles (e.g. Faber et al. 1997) and theoretical simulations (e.g. 
Navarro, Frenk \& White 1997, Moore et al. 1998). Most simple, modern, analytic 
representations of galaxy mass distributions have density distributions of the
form $\rho \propto r^{-\gamma} (r+a)^{-n+\gamma}$ where $\gamma$ is the exponent
of the central density cusp, $n$ is the exponent for the asymptotic slope
of the density distribution, and $a$ is the break radius.  Popular limits
are the Hernquist (1990) model ($\gamma=1$, $n=4$), the Jaffe (1983) model
($\gamma=2$, $n=4$), the more general $\eta$-models (Dehnen 1993, Tremaine et 
al. 1994, $n=4$), the Navarro, Frenk \& White (1997, NFW) models ($\gamma=1$, 
$n=3$) and the more general Zhao (1996) models.  The classical de 
Vaucouleurs (1948) profile for early-type galaxies is similar to the
Hernquist model, although its central ``cusp'' is better approximated by a 
slightly shallower exponent (Hernquist 1990).

Few of these more realistic models have been considered as lens models.  Lehar et al.
(2000) fit SIE and ellipsoidal de Vaucouleurs models as standard models which 
represent the two typical limits of central cusps.  Only one analytic study by 
Evans \& Wilkinson (1998) explored the general mathematics of models with 
different central cusps, focusing on circular models because there are no general 
analytic results for cusped ellipsoidal models.  Bartelmann (1996)
explored using the NFW density distribution to explain the radial arcs 
sometimes found in cluster lenses, and Wyithe, Turner \& Spergel (2000) 
and Keeton \& Madau (2001)
explored the general statistical properties of a family of circular, cusped
models for clusters.  Cusps have qualitatively different lensing properties
from finite cores, making it important to begin exploring their effects on
gravitational lenses.
In \S2 we introduce a general class of ellipsoidal, cusped density distributions
for use in gravitational lensing, and in \S3 we apply the new models to the
gravitational lenses APM~08279+5255 and B~1933+503.  Our results are 
summarized in \S4.  The models are implemented and available as part of 
the {\it lensmodel} package of lens modeling codes 
(Keeton 2001).\footnote{The {\it lensmodel} package can be obtained from 
http://cfa-www.harvard.edu/castles}
   
\section{A New Class of Ellipsoidal Lens Model}

Our models are defined from the spherical density distribution,
\begin{equation}
    \rho = { \rho_0 \over r^\gamma } { a^n \over (r^2+a^2)^{(n-\gamma)/2} }
\end{equation}
for cusp exponent $\gamma$, break radius $a$ and outer exponent $n$ 
(it is the same as a Zhao  (1996) model with $\alpha=1/2$).  This
differs from the standard cusped models in using $(r^2+a^2)^{(n-\gamma)/2}$
rather than $(r+a)^{n-\gamma}$, leading to a smoother break in the density
profile, more convenient analytic properties for the lensing calculations,
and overlaps with several standard lens models.  This form also has physical
advantages in some cases.  For example, the rotation curve of the 
Jaffe (1983) model, $\rho \propto r^{-2}(r+a)^{-2}$, declines linearly
for $r<a$ despite the $1/r^2$ central density profile, while our
``pseudo-Jaffe'' model with $\rho \propto r^{-2}(r^2+a^2)^{-1}$ has
a fairly flat rotation curve for $r < a$.  Whenever the exponents of our
density model match those of a standard cusped model we will refer to
it as a ``pseudo-'' model.  Hence for $n=4$, we have the pseudo-Jaffe
model ($\gamma=2$), the pseudo-Hernquist model ($\gamma=1$), and the
pseudo-$\eta$ models (any $\gamma$), and for $n=3$ and $\gamma=1$
we have the pseudo-NFW model.  The model also matches the 
standard SIE model ($\gamma=2$, $n=4$ and $r \ll a$) and can
match all the softened power-law models (for $\gamma=0$ and 
viewing the break radius $a$ as a core radius).

\begin{figure}
\centerline{\psfig{figure=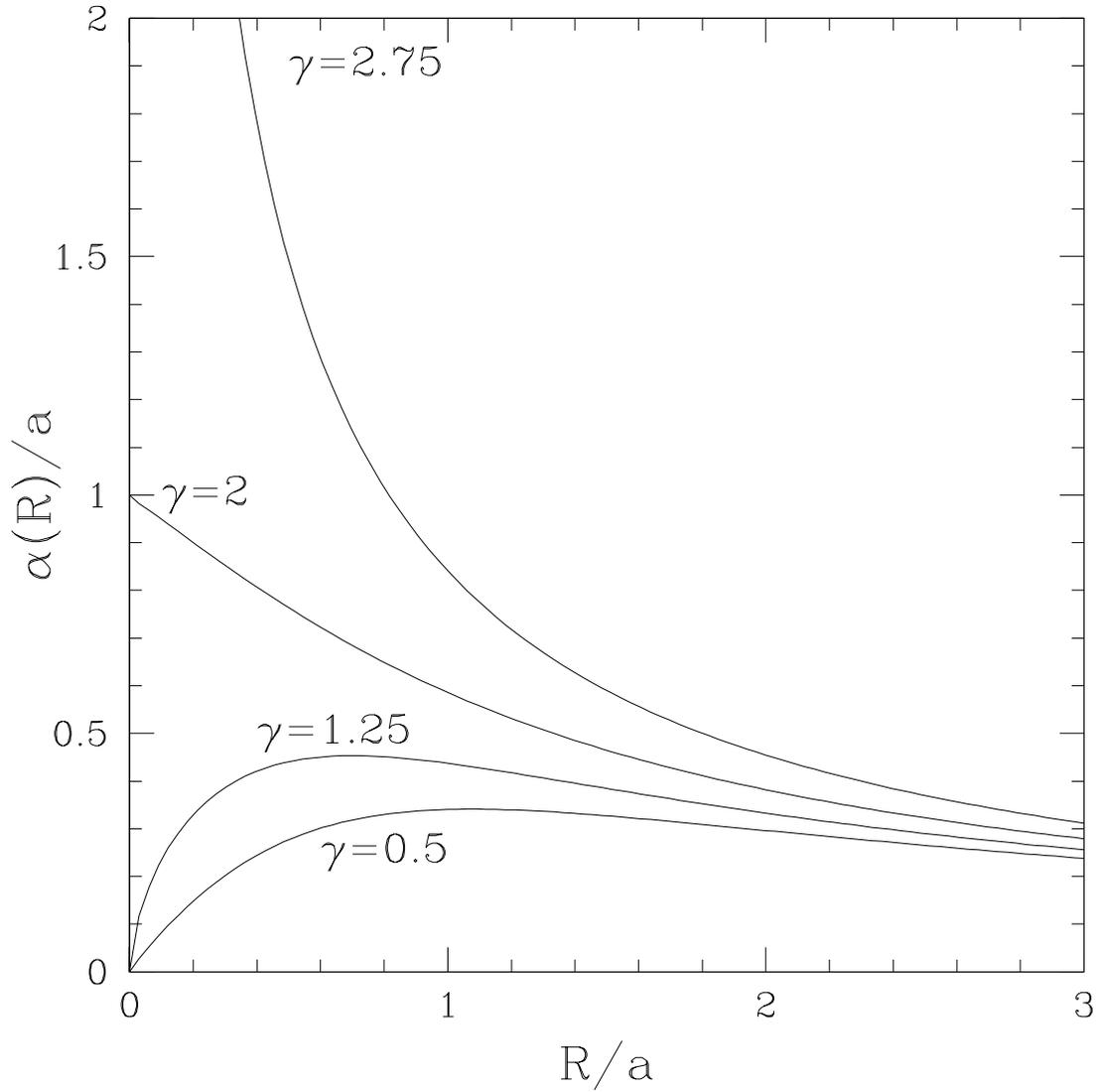,width=6in}}
\figcaption{The deflection angles of cusped models for several values of 
  $\gamma$, where the central density $\rho \propto r^{-\gamma}$, for a 
  fixed outer exponent of $n=4$ and a mass normalization of 
  $M_T/\pi a^2 \Sigma_{crit}=1$. 
       }
\end{figure}

Given the density distribution, the total mass of the model is
\begin{equation}
M_T=2 \pi a^3 \rho_0 B\left(\frac{n-3}{2},\frac{3-\gamma}{2}\right),
\end{equation}
where $B(a,b)=\Gamma(a)\Gamma(b)/\Gamma(a+b)$ is the Euler beta function.
The mass is well-defined (i.e. finite) only for $\gamma < 3$ and $n>3$.
The surface mass density in units of the critical surface density for
lensing is
\begin{equation}
\kappa(R)=\frac{\Sigma(R)}{\Sigma_{crit}}=
      \kappa_0\,B\left(\frac{n-1}{2},\frac{1}{2}\right)\,(1+\xi^2)^{(1-n)/2}\,
             {_2}F_1\left(\frac{n-1}{2},\frac{\gamma}{2},\frac{n}{2}; \frac{1}{1+\xi^2}\right)
\end{equation}
where the normalization $\kappa_0=\rho_0\,a/\Sigma_{crit}$, ${_2}F_1$ is the 
hypergeometric function  and $\xi=R/a$ is the projected radius $R$ normalized by the
break radius $a$.  For a circular lens with $n \not= 3$, the deflection angle is
\begin{equation}
\alpha(R)=\frac{2\, \kappa_0\,a}{\xi} \left[ B\left(\frac{n-3}{2},\frac{3-\gamma}{2}\right)
            -B\left(\frac{n-3}{2},\frac{3}{2}\right)\,(1+\xi^2)^{(3-n)/2}\,
            {_2}F_1\left(\frac{n-3}{2},\frac{\gamma}{2},\frac{n}{2}; \frac{1}{1+\xi^2}\right)\right]
\end{equation}
For $n=3$ the deflection must be expressed in terms of generalized hypergeometric functions.
The first derivative of $\kappa$, which is used in some of the integrals needed to compute
the properties of ellipsoidal models, is 
\begin{eqnarray}
\frac{d\kappa(R)}{dR} & = & -\frac{\kappa_0}{a}\,B\left(\frac{n+1}{2},\frac{1}{2}\right)
                           \xi\,(1+\xi^2)^{-(n+3)/2}  \left[\,n\,(1+\xi^2)\,\,\,
 {_2}F_1\left(\frac{n-1}{2},\frac{\gamma}{2}, \frac{n}{2}; \frac{1}{1+\xi^2}\right)\right. +
 \nonumber \\
&   &  + \left.  \gamma\,\,\,
{_2}F_1\left(\frac{n+1}{2},\frac{\gamma+2}{2}, \frac{n+2}{2}; \frac{1}{1+\xi^2}\right)\right]
\end{eqnarray}
In Figure 1 we plot the deflection angle, $\alpha(R)$, for several values of $\gamma$, a 
fixed outer exponent of $n=4$ and a fixed surface density scale 
$M_T/\pi a^2 \Sigma_{crit}=1$. The central deflection diverges for 
$\gamma > 2$, is finite for $\gamma=2$ and passes through zero for 
$\gamma < 2$.

The properties of the lens models in the core can be better understood by examining power
series expansions of the deflection angle and the magnification as $\xi=R/a \rightarrow 0$.
If we define the constants $A\equiv \kappa_0\,B[(n-1)/2,(1-\gamma)/2]$ and 
$B\equiv \kappa_0\,B[(\gamma-1)/2,1/2]$, for $\gamma\neq 1$ we find that
\begin{equation}\label{ks}
 \kappa(R)=A\,+\,B\,\,\xi^{1-\gamma}\,+\,O(\xi^2,\xi^{3-\gamma})
\end{equation}
\begin{equation}\label{as}
 \alpha(R)=R\,\left[A\,+\,\frac{2\,B}{3-\gamma}\,\,\xi^{1-\gamma}\,+\,O(\xi^2,\xi^{3-\gamma})\right]
\end{equation}
with logarithmic divergences when $\gamma=1$. For circular lenses, the inverse of the magnification,
$\mu^{-1}=1-2\kappa(R)-\alpha^2(R)/R^2+2\kappa(R)\alpha(R)/R$, can be expressed in 
terms of $\kappa(R)$ and $\alpha(R)$ (e.g. Schneider, Ehlers and Falco 1992). 
Using eqns. (\ref{ks}) and (\ref{as}), we find a power series expansion for the
central magnification of 
\begin{equation}
\mu^{-1}=(A-1)\,\,(A-1+2\,B\,\,\xi^{1-\gamma})+O(\xi^{2(1-\gamma)},\xi^2,\xi^{3-\gamma}).
\end{equation}
For $\gamma\ge2$ the cusp model produces only an even number of images; there is no central
``odd'' image. For $\gamma<2$ the model does produce a central, odd image, but for
$1<\gamma<2$ the image can be significantly demagnified compared to the other images
(e.g. Evans \& Wilkinson 1998).

When we make the models ellipsoidal we must calculate the deflections and magnifications 
numerically.  The projected radius $R$ is replaced by the ellipsoidal coordinate 
$R^2\rightarrow \eta^2=x^2+y^2/q^2$ where $q$ is the projected axis ratio and 
the Cartesian coordinates $x$ and $y$ are rotated to match the principal axes
of the lens.  The elliptical symmetry reduces the computation of the lensing properties to one-dimensional
integrals (see e.g. Schramm 1990). Modeling lenses using the numerical integrals is too slow for
practical use, so we tabulate the potential, the deflections and the components
of the magnification tensor as a function of scaled radius ($x/a$ and $y/a$) 
and axis ratio ($q$).  The modeling is then done
by interpolating over the tabulated models.  We have implemented
the ellipsoidal cusped models as part of the {\it lensmodel} software package (Keeton 2001).

\section{Application of the New Model to APM~08279+5255 and B~1933+503}

We now apply these models to two gravitational lenses.  The first lens,
APM~08279+5255 (Irwin et al. 1998), is the first clear example of a 
gravitational lens with a bright central image (see Ibata et al. 1999, 
Egami et al. 2000).  In this system we can explore the role of a shallow 
central cusp, rather than a finite core radius, in producing a visible 
central image.  The second lens, B~1933+507 (Sykes et al. 1998), is a 10 
image gravitational lens which is well-suited to determining the radial 
mass distribution of the lens galaxy.  A detailed analysis by 
Cohn et al. (2001) found that models with cusps slightly shallower than 
isothermal produced the best fits, but these models were not globally
consistent due to their infinite masses and perpetually rising rotation curves.

\subsection{APM~08279+5255}

When Irwin et al. (1998) discovered the $z_s=3.87$ BAL quasar APM~08279+5255 they
also noted that the image was extended and likely to be a compact double (confirmed
by Ledoux et al. 1998).  HST images in the infrared with NICMOS (Ibata et al. 1999)
and improved ground-based images (Egami et al. 2000) demonstrated that the source
was actually a triple. There are two bright point sources separated by $0\farcs38$ 
with a flux ratio of 0.8, and between the two lies a relatively bright third image with
a flux ratio of 0.2 relative to the brightest image. A lens galaxy has
yet to be detected, although this is not surprising given the extraordinary
brightness of the quasar images (about 15~R~mag!).  The image configuration
is easily modeled as an isothermal ellipsoid with a finite core
radius (Ibata et al. 1999, Egami et al. 2000).  However, given our expectation 
that the lens galaxy should not have a core radius, we will investigate 
whether the system is equally well modeled by a shallow cusp.

\begin{figure}
\centerline{\psfig{figure=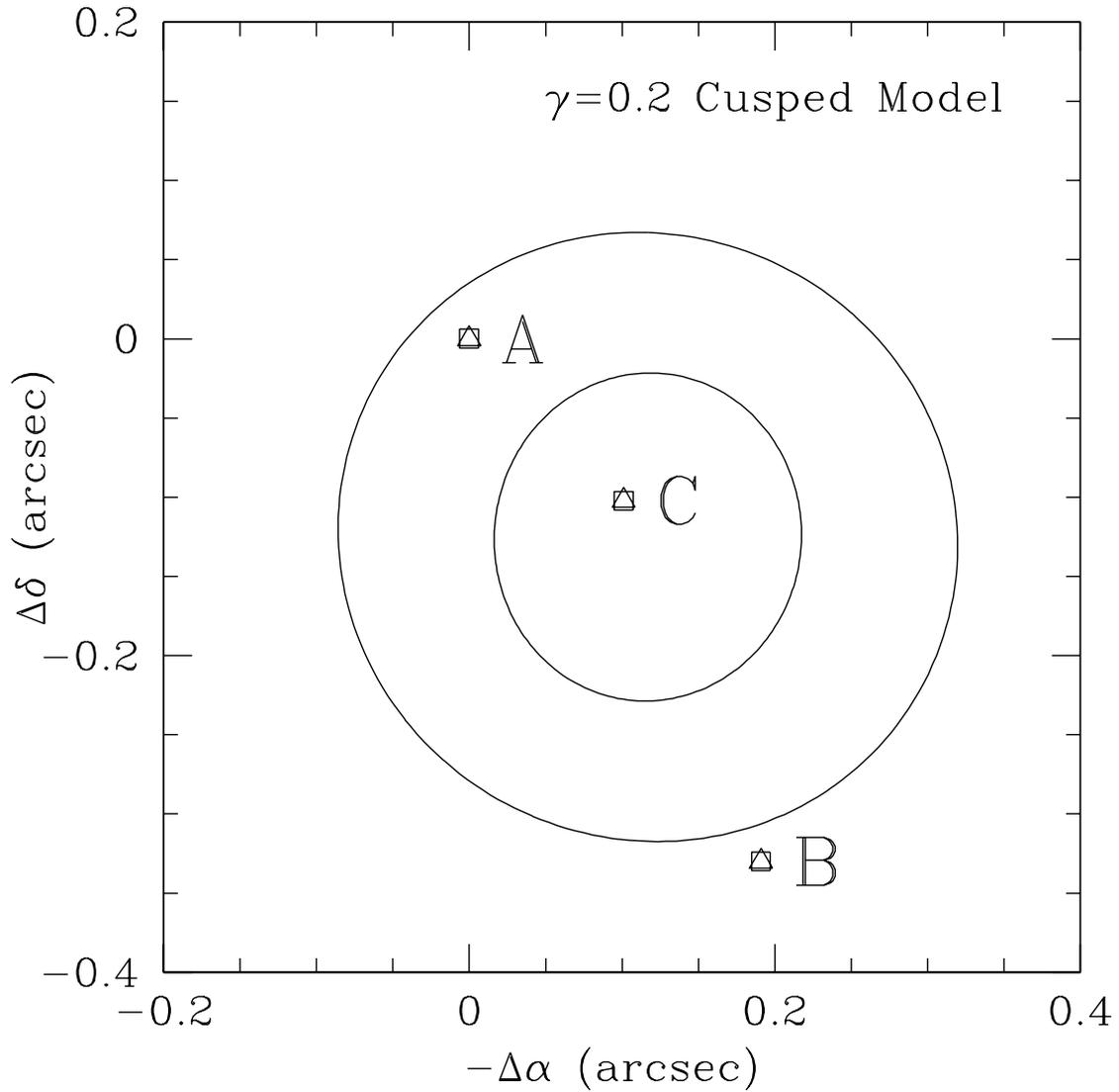,width=6in}}
\figcaption{A schematic diagram of APM~08279+5255. The squares show the observed positions of the three quasar
  images, and the triangles show the model positions in the best fitting $\gamma=0.2$ cusped model from 
  Table 2.  The inner and outer curves are the radial and tangential critical lines of the model.
  }
\end{figure}

\begin{figure}
\centerline{\psfig{figure=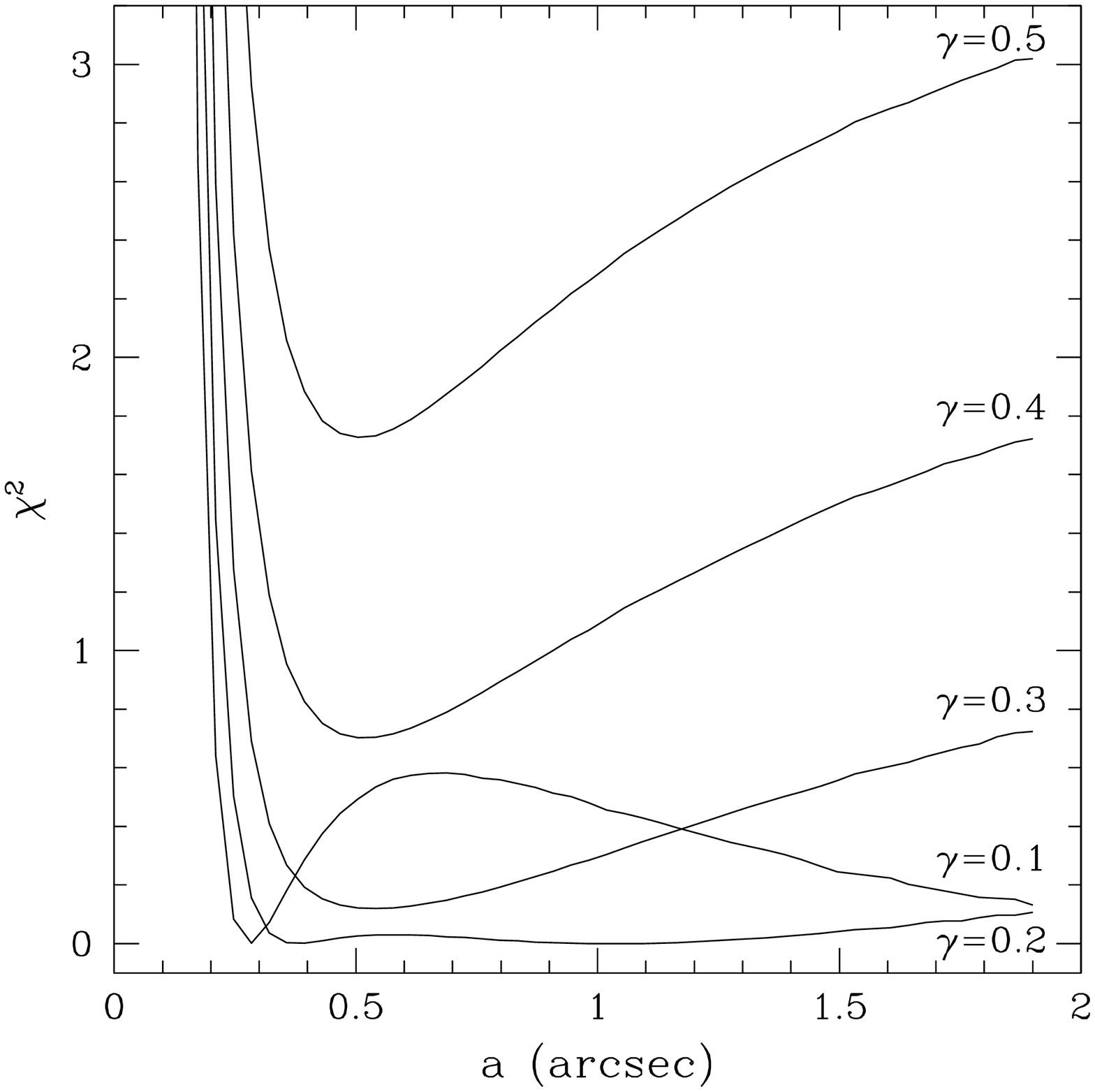,width=6in}}
\figcaption{ Ellipsoidal cusped models for APM~08279+5255. We show the $\chi^2$ statistic for the
 lens models as a function of the break radius $a$ for several values of the central cusp 
 exponent $\gamma$.   Note the presence of high (large $a$) and low (small $a$) magnification
 solution branches for $\gamma=$0.1 and 0.2. 
  }
\end{figure}

While it is likely that the system is composed of three lensed images 
given the fluxes and the similarity of the infrared colors (see Ibata et al. 
1999), we can use optical flux ratios for further confirmation.  The 
CfA/Arizona Space Telescope Lens Survey (CASTLES) obtained I-band (F814W)
and V-band (F505W) images of APM~08279+5255 on 1999.11.09 with the Hubble
Space Telescope (HST).  Each observation consisted of 20 dithered 40~sec exposures 
designed to avoid saturating the extraordinarily bright quasars.  The data 
were reduced and analyzed following the standard procedures of 
Lehar et al. (2000) for the CASTLES observations.
Table 1 presents the resulting astrometry and fluxes for the system
where we also include the photometry obtained from Ibata et al. (1999) with
the NICMOS infrared images.
We now see that not only are the infrared colors of the images similar, but the 
spectrum remains similar over the full range from the V-band to K-band, 
which makes it virtually impossible for the three images to be anything 
other than three images of the BAL quasar.  We use these results to
constrain our lens models with a more conservative  error of 0\farcs01 for the quasar positions.

We first fit the system using a softened isothermal ellipsoid. Like 
Ibata et al. (1999) and Egami et al. (2000) we find that the model produces a 
perfect fit to the data given a core radius of $a=0\farcs16^{+0\farcs10}_{-0\farcs05}$, which would 
be $0.7^{+0.4}_{-0.2}\,\,(\,0.9^{+0.6}_{-0.3}\,)\,\, h^{-1}$~kpc for a lens redshift of $z_l=0.5$ ($1.0$) in an 
$\Omega_0=0.3$ flat cosmological model.  Since the bulges of massive galaxies
usually show cusps down to smaller scales (see e.g. Faber et al. 1997), this solution appears to 
be inconsistent with the known properties of galaxies.  The best model reaches 
$\chi^2=0$, as we would expect given a system with $6$ constraints and $6$ 
parameters leaving $N_{dof}=0$ degrees of freedom.  
We also fit a model where the lens was an exponential disk
without a bulge or dark matter halo.  Exponential disks have little density
contrast between their centers and the Einstein ring, which makes it relatively
easy to fit the system.  We show the parameters of these two models in Table 2.
Lens models with a de Vaucouleurs profile were not able to fit the data for any effective radius.
The central density of the $R^{1/4}$ law roughly corresponds to a central density 
distribution with  $\rho\sim r^{-3/4}$ as $r\rightarrow 0$ (see Hernquist 1990), 
and a central exponent $\gamma=3/4$ is too steep to reproduce the central bright 
image (see below).

Next we fit a sequence of our cusped models to the system (see Figures 2 \& 3).
The models are underconstrained when we allow the
cusp exponent and the break radius both to vary ($N_{dof}=-1$
given 6 constraints and 7 parameters). Nevertheless, we could only
find successful models for cusps outside the standard range of
$1 \lesssim \gamma \lesssim 2$ found for stellar luminosity profiles
and dark matter simulations.
We found that the cusp exponent has to satisfy 
$\gamma \ltorder 0.4$ to achieve $\chi^2 <1$, and the models can produce
perfect fits only for $\gamma \ltorder 0.2$.  These cusps are so shallow
that we are almost at the limit of reproducing the softened power law 
models ($\gamma=0$).  While the cusp models with a small break radius
have total magnifications comparable to the softened isothermal models
($\mu_T \simeq 100$), there is a second solution branch with a very large
break radius where the total magnification slowly diverges.   
For example a model with $\gamma=0.2$ and $a\sim 1\arcsec$ (see
in Fig. 3) provides a perfect fit combined with a total magnification
of $\mu_T \sim 1800$.  Without any correction for lensing, the implied
luminosity of the system is $\sim 5\times 10^{15} L_\odot$ ($\Omega=1$, 
Egami et al.~2000), and in the previous models and most of our new 
models the lens magnification of $\mu_T \sim 100$ reduces this to a more 
physically plausible $\sim 5 \times 10^{13}L_\odot$. On the branch of solutions
with $\mu_T \sim 1800$, however, the intrinsic luminosity of the system
drops to $\sim 3 \times 10^{12}L_\odot$ and 
the source ceases to be remarkable in its intrinsic properties.   
We can find models with steeper central cusps ($1 < \gamma < 2$) by adding an 
external shear, but these models are very underconstrained (9 parameters for
6 constraints) and require very high lens ellipticities.  They are probably
unphysical, and we do not present any examples.  

\begin{figure}
\centerline{\psfig{figure=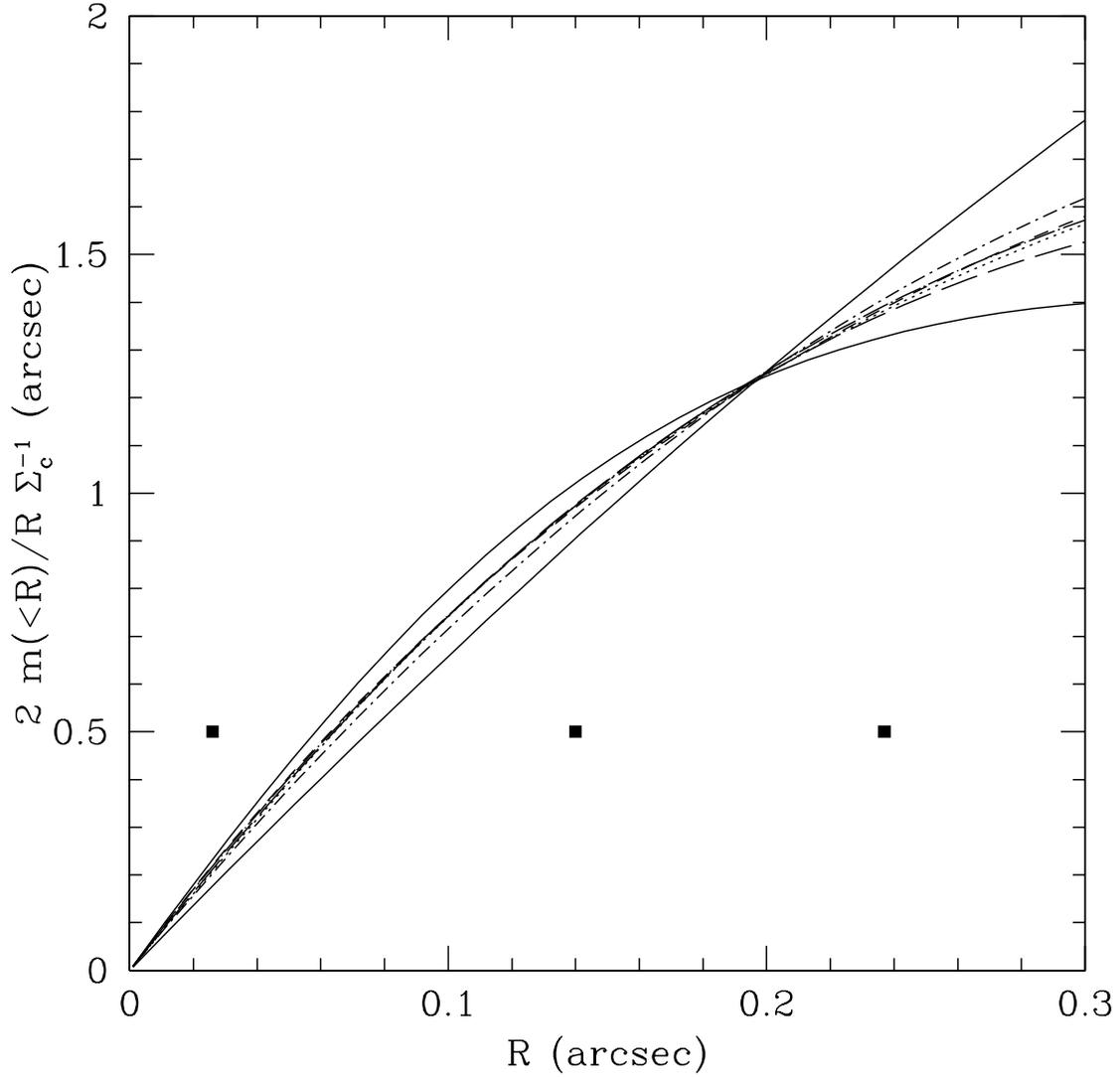,width=6in}}
\figcaption{Monopole deflections for the lens models of APM~08279+5255 listed in Table 2. At
r=0\farcs3 the smaller (larger) value corresponds to the $\gamma=0.1$ ($\gamma=0.2, a=1\farcs02$) 
cusped model. The rest of the models (SIE with a finite core (dotted), exponential disk (short dashed)
and $\gamma$=0.2 (long dashed), 0.3 (dot-short dashed), 0.4 (dot-long dashed) cusped models) are very 
similar at the image positions (marked by 
the solid points). In particular the softened SIE, exponential disk and $\gamma=0.4$ cusped models
are almost indistinguishable.
 }
\end{figure}

Table 2 summarizes a representative set of models.  Since the models are 
underconstrained and do not include an external shear (which we know is
required in a complete model, see Keeton, Kochanek \& Seljak 1997), we do not include
error estimates for the parameters.  It is impossible to distinguish between 
the models using the current constraints, in large part because the
actual deflection profiles of the models are very similar (see Fig. 4).
The position of the lens galaxy is predicted to vary by only a few
milli-arcseconds between the models, and it will be extraordinarily
difficult to detect the lens galaxy at all given the enormous flux
from the lensed images.  With the exception of the high magnification
models, the time delays vary by only 20\% and a precise knowledge of
the Hubble constant would be required to distinguish between the 
models. The delay ratios,
which are independent of the Hubble constant, vary by only 1\% between
the models.  The only prospect for significantly improving the constraints
in this system is to search for the lensed images of the quasar host 
galaxy (see Kochanek, Keeton \& McLeod 2001), although the brightness of the quasar
images makes this a challenging observation.
  
\subsection{B~1933+503}

B~1933+507 (Sykes et al. 1998, Nair 1998, Marlow et al. 1999, Browne et al. 
1999, Biggs et al. 2000, Norbury et al. 2000) has 10 lensed images of a 
$z_s=2.62$ radio source generated by a $z_l=0.76$ lens galaxy.  The source 
has three radio components, two of which have 4 lensed images and one of which
has two lensed images (see Fig. 5).  The large number of constraints provided 
by the 10 images
makes this lens an ideal laboratory for studying the radial mass distribution of the lens galaxy.
Extensive models by Cohn et al. (2001) found that the best mass distribution they considered was
a power law model with a singular core and an exponent slightly shallower than isothermal ($s=0$
and $\alpha \simeq 1.1$). The $\alpha \gtorder 1$ power-law models with zero core radius favored by the
Cohn et al. (2001) analysis correspond to our cusped models with $\gamma=3-\alpha$ in the limit
of infinite break radius, $a \rightarrow \infty$.  The Cohn et al. (2001) models are unphysical 
because they have infinite
mass and a rising rotation curve.  Our new cusp models allow us to examine these solutions using the 
same central cusp exponents combined with an outer break radius to produce a physical model.
We used the same constraints as Cohn et al. (2001) to study the system.

\begin{figure}
\centerline{\psfig{figure=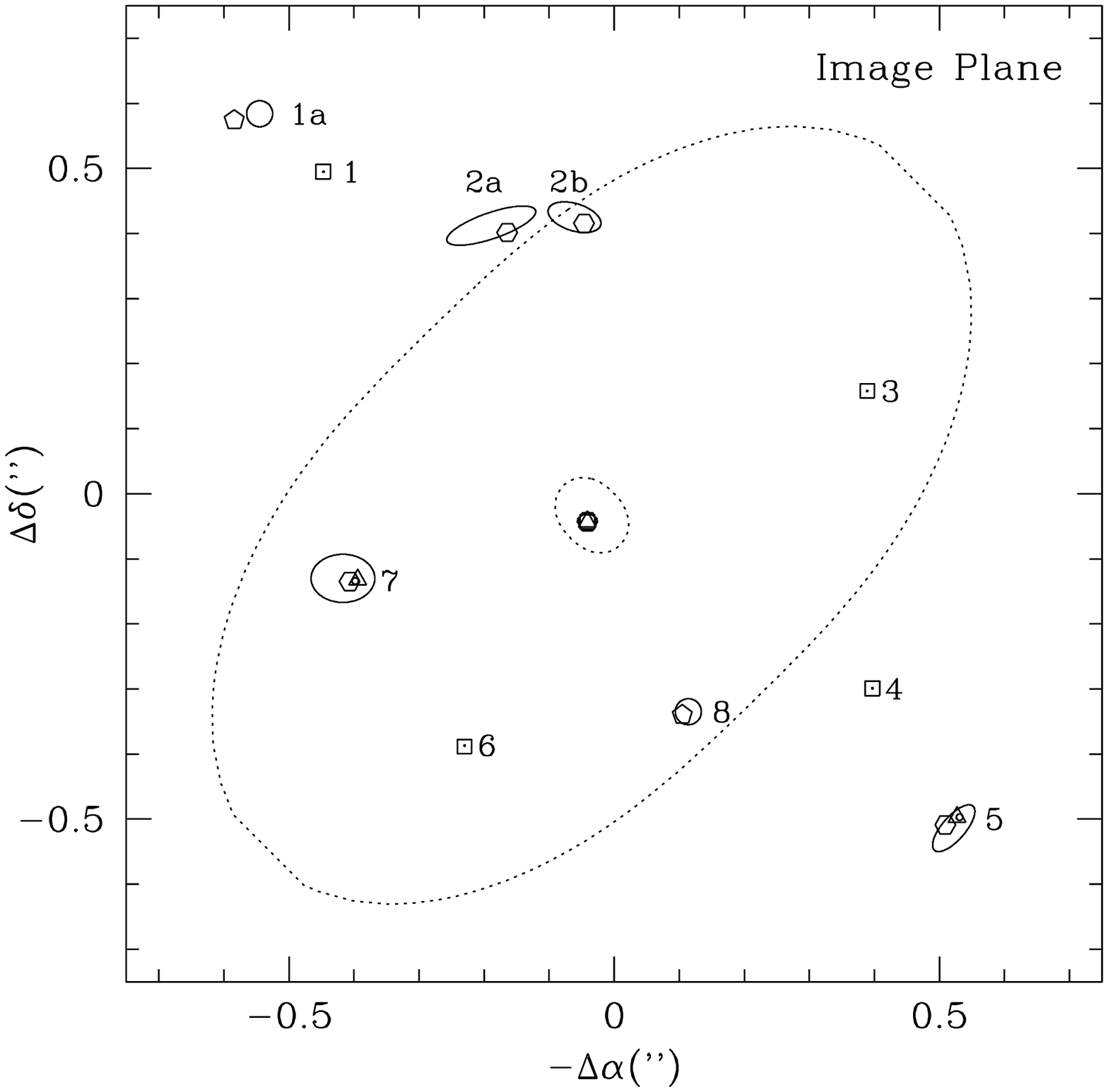,width=3.5in}
\psfig{figure=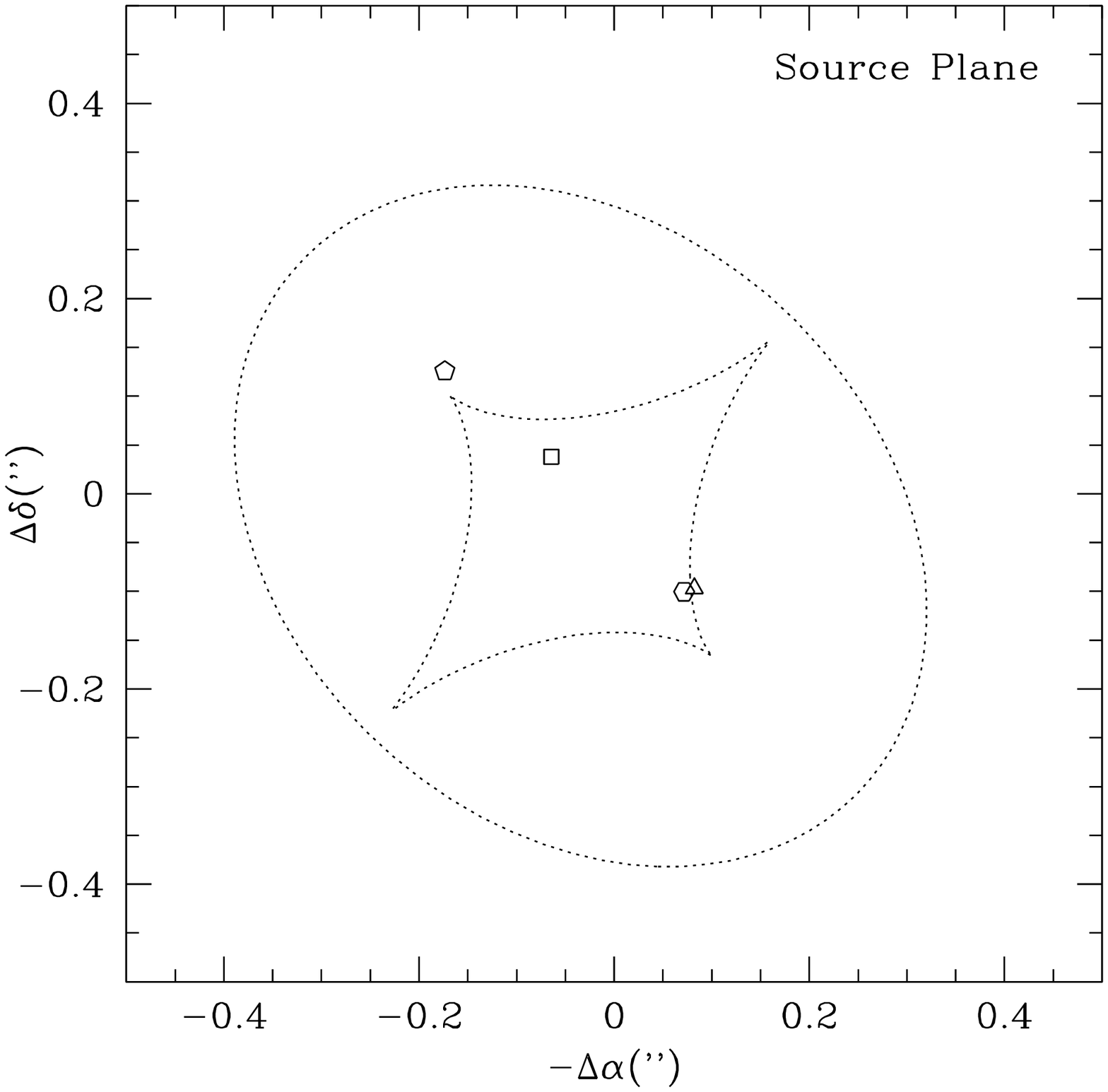,width=3.5in}}
\figcaption{Schematic diagrams of the image (left) and source (right) planes of B~1933+503.  The lensed image
  is composed of 3 source components. The compact core is lensed into 4 images (1, 3, 4 and 6) marked with
  squares in the left panel. The first steeper spectrum source component is lensed into 2 images (1a, 8) marked with
  pentagons.  The second steeper spectrum source component straddles a caustic and is lensed into a combination
  of two images (5 and 7) marked with triangles and four images (2a, 2b, 5 and  7) marked with hexagons.  The
  error ellipses on the image plane show the observed image positions and their uncertainties.  The points are
  the model image and source positions for the $\gamma=1.85$ cusp model from Table 2.   The inner (outer) 
  dashed lines on the image plane show the radial and tangential critical lines, while the outer (inner)
  dashed lines on the source plane show the radial and tangential caustics.
  }
\end{figure}

\begin{figure}
\centerline{\psfig{figure=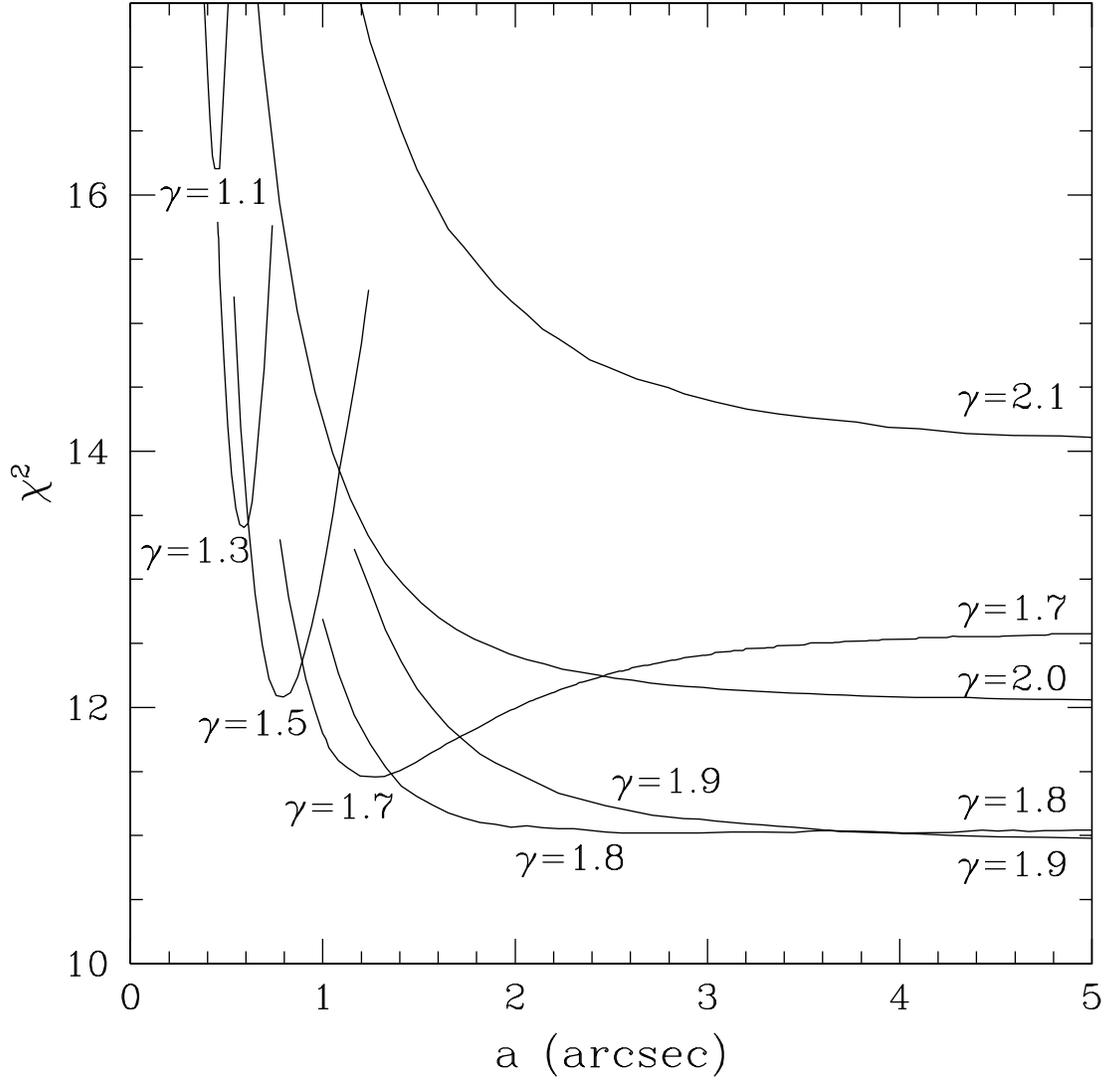,width=6in}}
\figcaption{Model $\chi^2$ statistics for B~1933+503 as a function of the break radius $a$ for a range 
 of cusp exponents, $\gamma$.  The lens model, consisting of an ellipsoidal cusp in an external shear field,
 has $N_{dof}=17$ degrees of freedom.  
 }
\end{figure}

\begin{figure}
\centerline{\psfig{figure=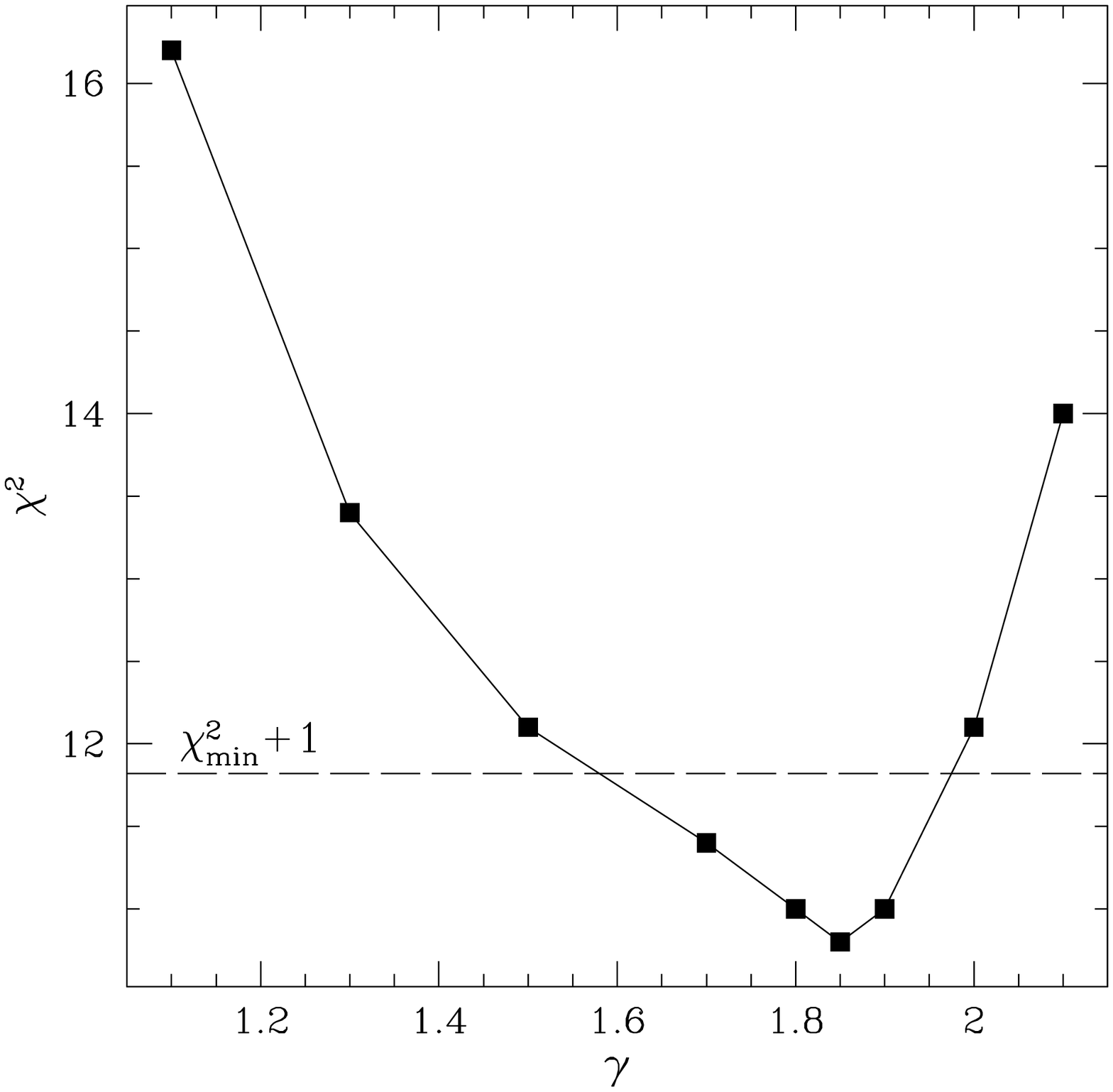,width=3.5in}
\psfig{figure=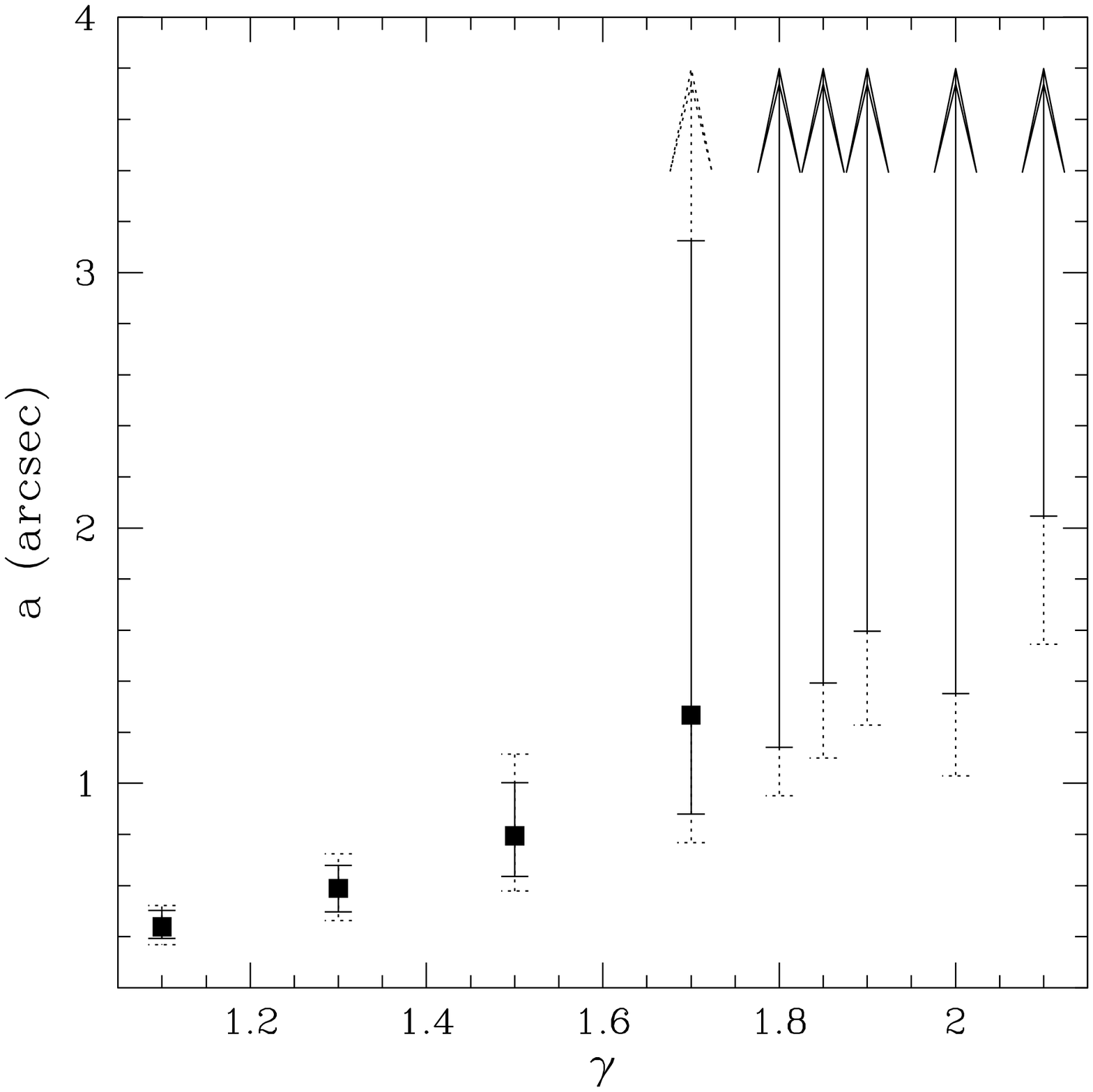,width=3.5in}}
\figcaption{(Left) The minimum $\chi^2$ statistic for B~1933+503 as a function of the cusp exponent 
  $\gamma$ after optimizing the break radius.
(Right) The optimal, 1-$\sigma$ and 2-$\sigma$ range limits in B~1933+503 for the break radius $a$ as a 
function of the cusp exponent $\gamma$.  The limits were determined by the $\Delta\chi^2=1$
range (solid lines) and the $\Delta\chi^2=2$ (dashed lines) from the minimum $\chi^2$ (points) 
found for each value of the cusp exponent. For $\gamma\ge 1.8$ the minimum $\chi^2$ is reached
when $a\rightarrow \infty$.}
\end{figure}

\begin{figure}
\centerline{\psfig{figure=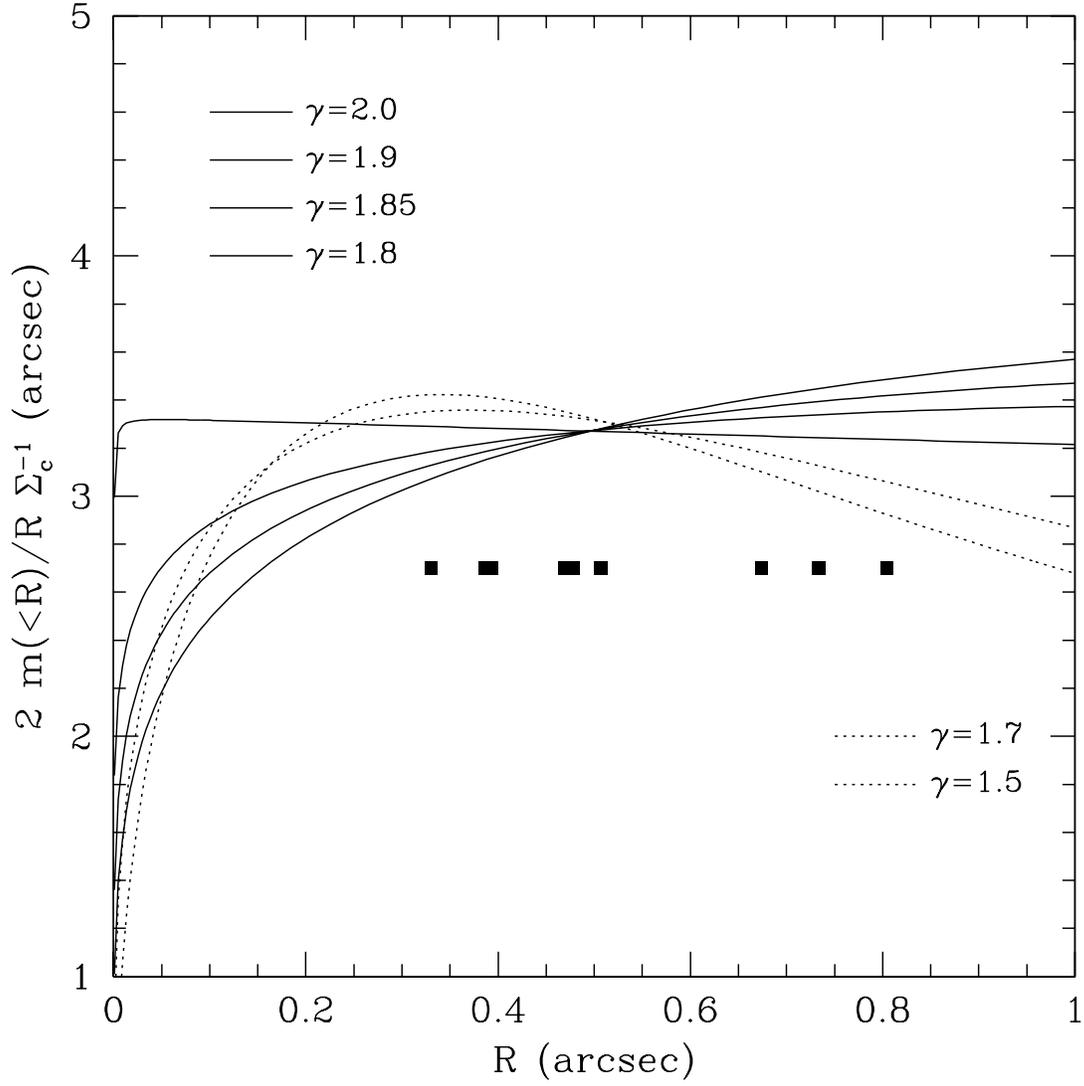,width=6in}}
\figcaption{The monopole deflection of the cusped models for B~1933+503. The solid (dotted) lines correspond at
  $r=0$ ($1\arcsec$) from bottom to top with cusp exponents $\gamma$=1.8,1.85,1.9,2.0 (1.5,1.7).
  The monopole deflection is roughly proportional to the square of the rotation curve, $v_c(r)^2$. 
  The points show the radial positions of the lensed images.
  }
\end{figure}

\begin{figure}
\centerline{\psfig{figure=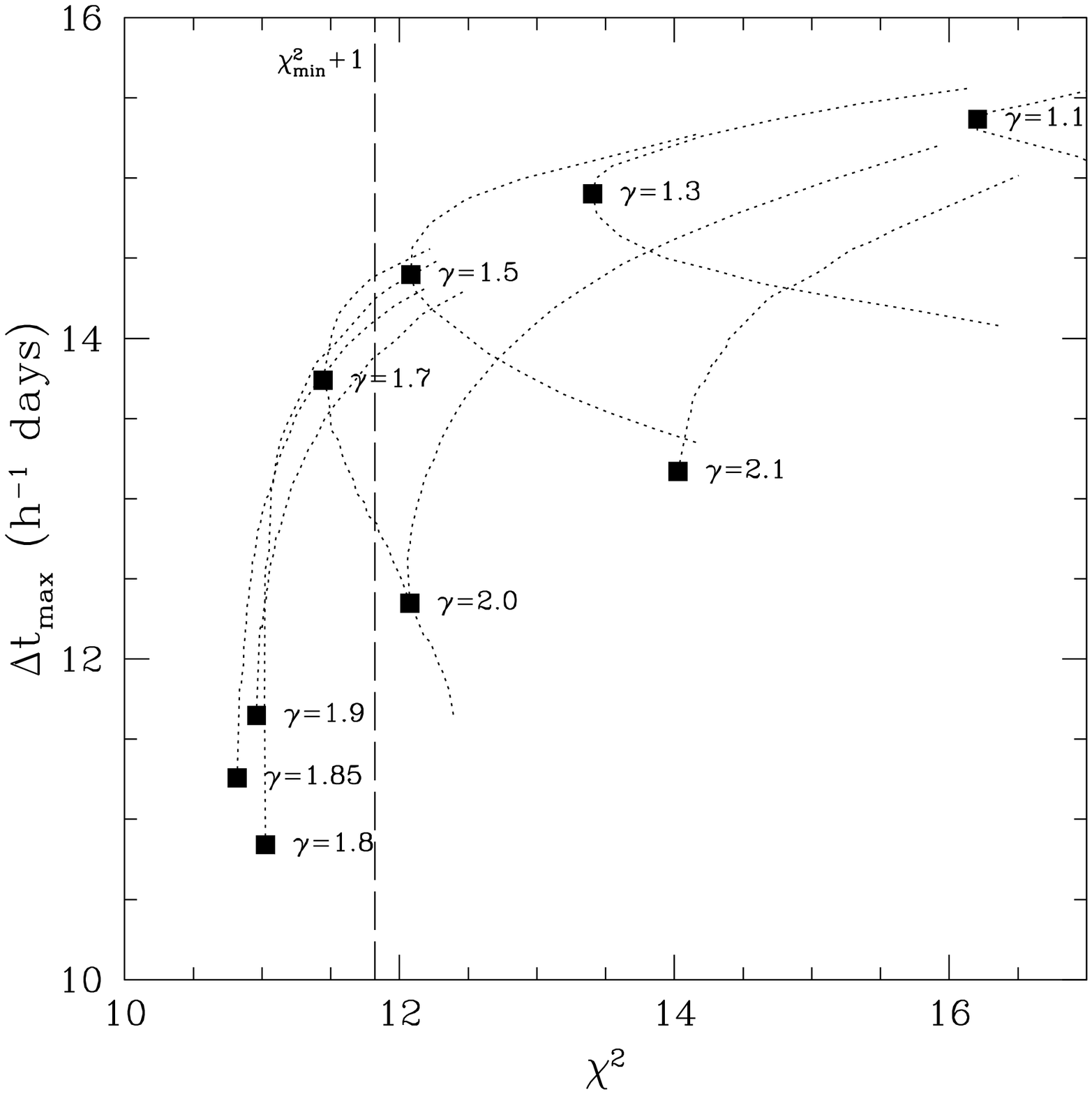,width=6in}}
\figcaption{Time delay predictions for B~1933+503 as a function of the $\chi^2$.
We display only the longest time delay between the four images of the
compact radio-core (it corresponds to images 1 and 6, see Figure 5) given in days in a flat
($\Omega_0=0.3$) cosmology with $H_0=100 h$~km~s$^{-1}$~Mpc$^{-1}$.
The points correspond to the best lens models for different values of $\gamma$. The dotted lines show
the variation of the time delay as a function of the $\chi^2$ when we vary the break
radius for each value of $\gamma$. The dashed line shows the limit $\Delta\chi^2\le1$.
For $\gamma \geq 1.8$ the solid points correspond to break radius of 10\farcs0, and further
increases in the break radius do not significantly reduce the delays (for $\gamma=2$, the
delay decreases by only 2.5\% as the break radius rises from $a=10\farcs0$ to $a=\infty$).} 
\end{figure}

Figure 6 shows the goodness of fit as a function of the break radius for a representative range 
($1.1\leq\gamma\leq 2.1$) of cusp exponents. Cusps shallower than $\gamma = 1.1$ are ruled out 
because they produce observable central images.  The best fit models for a range of cusp 
exponents are presented in Table 3, and Figure 7 shows goodness of fit and the permitted
range of break radii as a function of the cusp exponent.
The best fit model has $\gamma=1.85$ and a large break radius (which we fixed to $a=10\arcsec$)
with a total fit statistic $\chi^2=10.8$ for 17 degrees of freedom given that the cusped model
has 9 parameters and the data supplies 26 constraints (see Cohn et al. 2001).  Figure 5
compares the data and the model images and shows the locations of the source
components relative to the model caustics.  The largest contribution to the $\chi^2$ (4.3) 
is due to the position of image 1a (see Fig. 5).  This best fit model is nearly identical
to that of Cohn et al. (2001) who considered models with the same central density exponent
but an infinite break radius.   

The cusp exponent is limited to the range $1.6 \ltorder \gamma \ltorder 2.0$, marginally 
excluding the isothermal exponent $\gamma=2$ and strongly excluding the 
popular $\gamma=1$ cusps of the NFW and Hernquist models.  Shallower cusps, 
$\gamma < 1.8$, require finite break radii that become steadily smaller 
as the cusp becomes shallower.  In other words, to maintain a large enough surface density 
contrast between the center of the lens and the Einstein ring to avoid 
creating visible central images, the shallower cusps require global profiles 
that rapidly break onto the steeper outer profile.  As pointed out by 
Rusin \& Ma (2000), shallow cusps of infinite extent ($\gamma < 1.8$ with 
$a\rightarrow \infty$) can be ruled out because they will produce visible 
central images in too many lensed systems to be consistent with the data.  
Here, we can rule out the shallower cusps for any break radius, rather than
the limiting case of a break radius which is large compared to the Einstein
ring of the lens.  Steeper central cusps, $\gamma > 2$, intrinsically lack 
central images (see \S2), but are also inconsistent with the constraints.  

Figure 8 shows the monopole deflection of lens models which are consistent 
with the data.  Despite the enormous freedom in the models, only models with 
nearly flat rotation curves are consistent with the data. 
The dashed and solid lines are used to distinguish between the models with
$\gamma<1.8$ where the best lens model has a finite break radius and $\gamma\ge1.8$
where the best model is reached for $a\rightarrow \infty$. This discontinuity
produces two different trends in the monopole deflection as well as a small shift
in the crossing point.

The location of the outer break radius does affect the time delays between the images, 
where for a given $\gamma$ we find shorter delays as we increase the break radius. 
Figure 9 shows the longest time delay between the four images of the compact radio-core 
(it corresponds to images 1 and 6, see Figure 5) as a function of the $\chi^2$. The points
correspond to the best lens models for different values of $\gamma$. The dotted lines show
the variation of the time delay as a function of the $\chi^2$ when we vary the break
radius for each value of $\gamma$. The dashed line shows the limit $\Delta\chi^2\le1$. 
Given the generality of the cusped model we can consider the values inside this
region as a very robust prediction of the time delay. This figure also shows more clearly
the consequences of the jump in the optimal break radii from finite to infinite at
$\gamma \simeq 1.8$.

\section{Summary}

Gravitational lenses are used both to estimate the mass distribution of 
galaxies (e.g. Kochanek et al. 1989, Kochanek 1991, Kochanek \& Narayan 1992, 
Wambsganss \& Paczynski 1994, Kochanek 1995, Chen et al. 1995, 
Wallington, Kochanek \& Narayan 1996, Ellithorpe, Kochanek, \& Hewitt 1996,
Chae et al. 1998, Schmidt et al. 1998, Kochanek et al. 2000, Cohn et al. 2001)
and the Hubble constant from measurements of the 
time delays between the lensed images (e.g. Grogin \& Narayan 1996,
Kundic et al. 1997, Schechter et al. 1997, Keeton \& Kochanek 1997, 
Impey et al. 1998, Chae 1999, Barkana et al. 1999, Koopmans \& Fassnacht 1999, 
Fassnacht et al. 1999, Keeton et al. 2000, Kochanek, Keeton \& McLeod 2001,
Oscoz et al. 2001).
Almost all these analyses use parametric models for the mass distribution of the lens.
Parametric models have three advantages:  first, all standard parametric
models represent physically self-consistent, dynamically stable
equilibrium systems;  second, parametric models are relatively easy to 
interpret physically; and third, they are computationally efficient.  The 
disadvantage of parametric models, particularly for estimating the
systematic uncertainties in the mass distribution or the Hubble 
constant, is that they may overly restrict the true freedom in the mass
distributions of real astronomical systems. 

Our new family of ellipsoidal models for gravitational lenses is a
significant advance for two reasons.  First, the new models are far
more realistic models of galaxies than the standard lens models given
our current understanding of the central stellar density profiles of
galaxies (e.g. Faber et al. 1997) and simulations of dark matter halos 
(Navarro et al. 1997, Moore et al. 1998), where the
central density follows a power-law cusp with $\rho \propto r^{-\gamma}$ 
and $1 \ltorder \gamma \ltorder 2$.  Our model, which is a subcase of
the more general Zhao (1996) models, includes analogues of many 
currently popular analytic density distributions such as the Jaffe
(1983) model, the Hernquist (1990) model, the $\eta$ model 
(Dehnen 1993, Tremain et al. 1994) and the NFW model (Navarro et al. 1997).  
Second, independent of the models being physically more realistic than
the standard lens models, they provide a genuinely different class
of density distributions from the standard lens models.  Since the
freedom in the radial mass distribution appears to be the dominant
source of systematic uncertainties in determining the Hubble constant
from gravitational lenses (see e.g  Impey et al. 1998,
Fassnacht et al. 1999, Keeton et al. 2000), the new models can be used as an
additional probe of the uncertainties.  The {\it cusped models} are 
implemented and publically available as part of the {\it lensmodel} 
(Keeton 2001) package for modeling gravitational lenses. 

Non-parametric models of the mass distribution avoid the restrictions of
parametric models.  In particular, the elegant linear programming method
introduced by Saha \& Williams (1997) assumes only that the surface mass
density of the lens is non-negative.  This method will find a broader
range of systematic uncertainties in the mass distribution or the Hubble
constant than any parametric model or set of parametric models.  
Unfortunately, the models also lose the first advantage of the parametric
models, that they correspond to physical, dynamically stable systems.  Physical
systems must have non-negative density distributions and phase space
distribution functions, and many non-negative surface densities 
will correspond to unphysical models with regions of negative density
or phase space density (see Cohn et al. 2001).  Models which are physical
may still be improbable because they are dynamically unstable, with
lifetimes of only a few dynamical (crossing) times.\footnote{It is
possible to detail these additional conditions on the surface 
density $\Sigma(R)$ for spherical systems (see Binney \& Tremaine
1987).  The mass density,
$\rho = -\pi^{-1} \int_r^\infty d R (d\Sigma/dR) (R^2-r^2)^{-1/2}$,
must be positive for all $r$, which is trivially satisfied if the 
surface density distribution is a monotonically declining function, 
$d\Sigma/d R < 0$.  For an isotropic distribution function, which
depends only on the binding energy $\epsilon$, the distribution 
function corresponding to a given density distribution is positive
if $\int_0^\epsilon (d\rho/d\psi) (\epsilon-\psi)^{-1/2} d\psi$ is
an increasing function of $\epsilon=\psi-v^2/2$.  This is trivially
satisfied if $d\rho/d\psi > 0$, which is true for all the parametric
models. The system is dynamically stable if the derivative of the 
distribution function $df/d\epsilon <0$.  Each step in this process
imposes a smoothness criterion (through the constraints
on the derivatives of the previous variable) which goes well beyond
simply requiring a positive surface density.  Viewed in terms of
the surface density $\Sigma$, a positive density is a restriction
on $d\Sigma/dR$, a positive distribution function is a restriction
on $d^2\Sigma/dR^2$, and stability is a restriction on 
$d^3\Sigma/dR^3$!  Unfortunately, there
are no analytic expressions even for spherical systems with 
distribution functions depending on energy and angular momentum,
let alone systems with more complicated symmetries.   
Although galaxies are more complicated than simple spherical
systems, the problems we illustrate using the spherical systems
apply to all galaxies and mean that a positive surface density
is only a first, minimal criterion for a physical mass model.  }
These additional
criteria mean that the current non-parametric methods greatly 
over-estimate the freedom in the mass distribution and frequently
produce unphysical models.  These issues make the two approaches to
modeling lenses complementary.  The parametric models are always 
physical but can underestimate the freedom in the mass distribution,
while the non-parametric models are frequently unphysical and 
overestimate the freedom in the mass distribution.

We explored these new parameterized mass distributions by using them to model two lenses, APM~08279+5255
and B~1933+503.  APM~08279+5255 (Irwin et al. 1998, Ibata et al. 1999, Egami et al. 2000) is the only
lens which clearly possesses three images of the background source.  We find further evidence of this
from the CASTLES V-band and I-band HST images of the system, which show that the flux ratios of the images 
change little from the K-band in the infrared to the V-band in the optical.  Although the ability of
the cusped models to model a three image gravitational lens like APM~08279+5255 without a finite core
radius was one of the motivations for exploring the cusped models, we actually found that APM~08279+5255
cannot be modeled by cusps in the favored range $1 \ltorder \gamma \ltorder 2$. 
Only models with
$\gamma \ltorder 0.4$, where the cusps become very similar to finite core radii
($\gamma=0$), could successfully 
explain the properties of this lens.  This suggests that the lens galaxy in APM~08279+5255 is a
spiral galaxy, where rotation curves are sometimes consistent with finite central densities rather
than steep central density cusps.  Alternatively, this could be an example of a ``disk''
geometry lens (Keeton \& Kochanek 1998) where we see three images produced by a nearly edge on
disk galaxy.  In a disk lens geometry, the lens galaxy should be offset perpendicular to the
A--B image separation.  The small image separation ($0\farcs38$) also argues
for a spiral lens (see Fukugita \& Turner 1991, Kochanek 1996), but the similar colors of the 
three images suggests that there is little dust along the three lines of sight through the lens.
Unfortunately, the extreme brightness of the lensed images makes it nearly impossible to find
the far fainter lens galaxy, let alone measure its position and morphology.  

The second lens we considered, B~1933+503 (Sykes et al. 1998), already had an 
extensive model survey suggesting that the best models would have steep central 
cusps (Cohn et al. 2001).  Here we show using physically realistic cusped
models rather than simple power law density distributions that the system 
requires a steep central cusp, $\rho \simeq r^{-1.8\pm0.2}$, with a break 
radius located well outside the Einstein ring of the lens.  Shallow cusps,
particularly the $\rho \propto r^{-1}$ cusps of the Hernquist and NFW models, 
are inconsistent with the data for any break radius.  Lenses with shallow
central cusps and break radii larger than the Einstein radius of the lens
generically predict visible central images, and are ruled out by the 
absence of such images in well-defined lens samples (Rusin \& Ma 2000).  
Shallow cusps combined with a break radius smaller than the Einstein ring
can demagnify the central image, but the monopole deflection curve
(the rotation curve) becomes centrally peaked and is ruled out by the
need for a flat deflection profile to fit the positions of the observed
images.
A massive central black hole can also demagnify central images (Mao, Witt \& Koopmans 2000), 
thereby weakening the limits on shallow cusps (or a central core radius) based on the
absence of central images.  In the case of B~1933+507, the restrictions on shallow cusps
are only partially due to the limits on the presence of central images, and when we
eliminate the flux limits on the presence of a central image the restriction on the
cusp exponent is only reduced to $\gamma \gtorder 1.5$ from $\gamma \gtorder 1.6$.
The generality of the new developed cusped models yields a very robust prediction of the time delay
for this system.  In particular the lower limit obtained in the time delay computation for
any cusped exponent $\gamma$ and any break radius (see Figure 9)
would produce a strong constraint in the upper value of the Hubble constant provided an observational
measurement of the time delay.

\bigskip
\noindent Acknowledgments:  We thank the CASTLES collaboration (E.E. Falco, C.D. Impey, 
C.R. Keeton, C.S. Kochanek, J. Lehar, B.A. McLeod, J.A. Mu\~noz \& H.-W. Rix) 
for permission to include their APM~08279+5255 
data.  We also thank Joanne Cohn for providing us with her constraints for B~1933+507 so that
we could exactly match her models of B~1933+503.  CSK is supported by the Smithsonian
Institution and NASA grant NAG5-9265.

%%TABLES%%%
\newpage
\begin{deluxetable}{ccccccccc}
\tablecaption{APM\,08279+5255 Photometry and Astrometry}
\tablewidth{0pt}
\tablehead{ ID & $\Delta\alpha$ & $\Delta\delta$ & Magnitude  & Color1 & Color2 & Color3 & Color4 \nl
            &\arcsec &\arcsec & V & V--I & V--J & V--H & V--K}
\startdata
A & $\equiv$0         &         $\equiv$0 & $16.96\pm0.03$ & $1.71\pm0.06$ & $3.51\pm0.03$ & $3.86\pm0.03$ & $4.72\pm0.04$  \nl
B & --0.199$\pm$0.003 & --0.327$\pm$0.003 & $17.74\pm0.03$ & $2.10\pm0.04$ & $3.99\pm0.03$ & $4.37\pm0.03$ & $5.23\pm0.03$  \nl
C & --0.100$\pm$0.003 & --0.105$\pm$0.003 & $18.70\pm0.07$ & $1.77\pm0.09$ & $3.46\pm0.07$ & $3.96\pm0.07$ & $4.67\pm0.14$  \nl
\enddata
\tablecomments{The infrared images (H, J and K bands) were obtained by Ibata et al. (1999)
with the NIC1 camera at the HST. The optical images were acquired by the
CASTLES collaboration.}
\end{deluxetable}

\begin{deluxetable}{lcccccccccc}
\tablecaption{APM~08279+5255 Lens Models}
\tablewidth{0pt}
\tablehead{
  Model
 &$\Delta\alpha$ (\arcsec)
 &$\Delta\delta$ (\arcsec)
 &$b^*$ (\arcsec)
 &$a^{\dag}$ (\arcsec)
 &$\epsilon$ (1--$b/a$)
 &$\theta_\epsilon (^{\circ})$
 &$h\;\Delta{}t_{AB}^\ddag$
 &$h\;\Delta{}t_{CB}^\ddag$
 &$\mu_T$
 &$\chi^2$
}
\startdata
% model              x (''east) y (''north) b  a('') elip  theta_e  delay1 delay2 mag chi^2
SIE                 &  --0.117 & --0.125 & 0.42 & 0.16 & 0.019 & 79 & 0.49 & 0.63 & 125 & 0 \nl
Exp. disk           &  --0.116 & --0.124 & 1.38 & 0.37 & 0.022 & 80 & 0.56 & 0.72 & 113 & 0 \nl
Cusp ($\gamma=0.1$) &  --0.117 & --0.126 & 0.82 & 0.28 & 0.029 & 78 & 0.74 & 0.96 & 62 &  0 \nl
Cusp ($\gamma=0.2$) &  --0.117 & --0.125 & 0.66 & 0.39 & 0.020 & 76 & 0.51 & 0.66 & 129 & 0 \nl
Cusp ($\gamma=0.2$) &  --0.117 & --0.125 & 0.54 & 1.02 & 0.005 & 73 & 0.13 & 0.17 & 1814 & 0 \nl
Cusp ($\gamma=0.3$) &  --0.117 & --0.123 & 0.55 & 0.54 & 0.017 & 74 & 0.42 & 0.52 & 210 & 0.1 \nl
Cusp ($\gamma=0.4$) &  --0.116 & --0.120 & 0.51 & 0.50 & 0.025 & 75 & 0.60 & 0.71 & 118 & 0.7 \nl
\enddata
\tablecomments{\\
$^*$ In the case of the cusped models it corresponds to the dimensionless normalization $\kappa_0$ used in the text.\\
$^\dag$ ``a'' is the core radius in the SIE model, the scale radius for the exponential disk, and the break radius in
the cusped model.  \\
$^\ddag$For the time delay computation we used $z_l=1.18$, corresponding to a strong MgII absorption system, for
the unknown redshift of the lens.  The time delay is given in days in a flat ($\Omega_0=0.3$) cosmology with
$H_0=100 h$~km~s$^{-1}$~Mpc$^{-1}$. }
\end{deluxetable}

\begin{deluxetable}{lcccccccccc}
\tablecaption{B~1933+503 Lens Models}
\tablewidth{0pt}
\tablehead{
  Model
 &$\Delta\alpha$ (\arcsec)
 &$\Delta\delta$ (\arcsec)
 &$k_0$ 
 &$a$ (\arcsec)
 &$\epsilon$ (1--$b/a$)
 &$\theta_\epsilon (^{\circ})$
 &$\gamma_s$ (shear)
 &$\theta_{\gamma_s} (^{\circ})$
 &$h\;\Delta{}t_{max}^\dag$
 &$\chi^2$
}
\startdata
% model              x (''east) y (''north) b ('') elip  theta_e  gamma theta_g  delay chi^2

Cusp ($\gamma=1.1$) &  0.033 & --0.029 & 0.926 & 0.44       & 0.331 & --41.6 & 0.174 & --47.6 & 15.4 & 16.2 \nl
Cusp ($\gamma=1.3$) &  0.035 & --0.031 & 0.569 & 0.59       & 0.349 & --41.7 & 0.147 & --48.3 & 15.0 & 13.4 \nl
Cusp ($\gamma=1.5$) &  0.035 & --0.032 & 0.343 & 0.79       & 0.363 & --41.5 & 0.125 & --49.3 & 14.4 & 12.1 \nl
Cusp ($\gamma=1.7$) &  0.035 & --0.033 & 0.168 & 1.27       & 0.388 & --42.1 & 0.095 & --50.5 & 13.7 & 11.4 \nl
Cusp ($\gamma=1.8$) &  0.035 & --0.036 & 0.021 & $\equiv10$ & 0.406 & --42.3 & 0.022 & --73.9 & 10.9 & 11.0 \nl
Cusp ($\gamma=1.85$)&  0.034 & --0.033 & 0.018 & $\equiv10$ & 0.407 & --42.4 & 0.031 & --63.6 & 11.3 & 10.8 \nl
Cusp ($\gamma=1.9$) &  0.034 & --0.032 & 0.015 & $\equiv10$ & 0.407 & --42.5 & 0.041 & --57.7 & 11.6 & 11.0 \nl
Cusp ($\gamma=2.0$) &  0.032 & --0.031 & 0.011 & $\equiv10$ & 0.413 & --42.7 & 0.061 & --52.6 & 12.3 & 12.1 \nl
Cusp ($\gamma=2.1$) &  0.031 & --0.030 & 0.008 & $\equiv10$ & 0.415 & --42.6 & 0.089 & --50.1 & 13.2 & 14.0 \nl

\enddata
\tablecomments{ The best models for $\gamma \gtorder 1.8$ have fit statistics that decline monotonically for
  increasing break radii $a$.  We show models with the break radius fixed to $a=10\arcsec$ for these cases.  
  The sudden jump in the break radius leads to the discontinuity in the parameter trends between the 
  $\gamma=1.7$ and $\gamma=1.8$ models.
\\$^\dag$Longest time delay between the four images of the compact radio-core (it corresponds to images
1 and 6, see figure 5) given in days in a flat ($\Omega_0=0.3$) cosmology with $H_0=100 h$~km~s$^{-1}$~Mpc$^{-1}$. }
\end{deluxetable}

\end{document}